\documentclass[3p,twocolumn]{elsarticle}

\usepackage{hyperref}

\journal{New Astronomy}

\biboptions{authoryear}

\def\astrobj#1{#1}

 \usepackage{color}
\newcommand\ja{\textcolor{black}}  
\begin{document}

\begin{frontmatter}

\title{A time-series \emph{VI} study of the variable stars of the \\ globular cluster NGC~6397\tnoteref{mytitlenote}}
\tnotetext[mytitlenote]{Based on observations at Estaci\'{o}n Astrof\'{\i}sica 
de Bosque Alegre, Argentina, and at Las~Campanas Observatory, Chile}

\author{J. A. Ahumada\corref{mycorrespondingauthor}}
\address{Observatorio Astron\'{o}mico, Universidad Nacional de
C\'{o}rdoba,  Laprida~854, X5000BGR, C\'{o}rdoba, Argentina}
\cortext[mycorrespondingauthor]{Corresponding author}
\ead{javier.ahumada@unc.edu.ar}

\author{A. Arellano Ferro}
\address{Universidad Nacional Aut\'onoma de M\'exico, Instituto de Astronom\'{\i}a, AP 70-264, CDMX 04510, M\'exico}
\ead{armando@astro.unam.mx}

\author{I. H. Bustos Fierro}
\address{Observatorio Astron\'{o}mico, Universidad Nacional de
C\'{o}rdoba,  Laprida~854, X5000BGR, C\'{o}rdoba, Argentina}
\ead{ivan.bustos.fierro@unc.edu.ar}

\author{C. L\'azaro}
\address{Departamento de Astrof\'{\i}sica, Universidad de La~Laguna, E-38206 La~Laguna, Tenerife, Spain, \emph{and} \\
Instituto de Astrof\'{\i}sica de Canarias (IAC), E-38205 La~Laguna, Tenerife, Spain}
\ead{clh@iac.es}

\author{M. A. Yepez}
\address{Universidad Nacional Aut\'onoma de M\'exico, Instituto de Astronom\'{\i}a, AP 70-264, CDMX 04510, M\'exico}
\ead{myepez@astro.unam.mx}

\author{K. P. Schr\"{o}der}
\address{Universidad de Guanajuato, Departamento de Astronom\'{\i}a, Guanajuato 36000, M\'{e}xico}
\ead{astrokp85@gmail.com}

\author{J. H. Calder\'on}
\address{Observatorio Astron\'{o}mico, Universidad Nacional de
C\'{o}rdoba,  Laprida~854, X5000BGR, C\'{o}rdoba, Argentina, \emph{and} \\
Consejo Nacional de Investigaciones Cient\'{\i}ficas y T\'{e}cnicas
(CONICET), Buenos Aires, Argentina}
\ead{jehumcal@gmail.com}


 

\begin{abstract}
We present a new time-series \emph{VI} CCD photometry  of the  
globular cluster NGC~6397, from which  we obtained and analysed the light 
curves of 35 variables carefully identified in the cluster field. We assessed the
 membership of the variables  with  an
astrometric analysis based on \emph{Gaia}~DR2 data. The cluster colour-magnitude diagram was differentially 
dereddened and cleaned of non members, which allowed us to fit isochrones
for [Fe/H]$ = -2.0$~dex in the range 13.0--13.5~Gyr, for a mean reddening $E(B-V)=0.19$, and  
a distance of 2.5~kpc. This distance was confirmed  using the period-luminosity
relation for the cluster's five SX~Phoenicis variables (V10, V11, V15, V21, and V23) 
present among its blue stragglers, yielding $2.24\pm0.13$~kpc.
We also modelled the light curves of four eclipsing binaries (V4, V5, V7, and
V8), and gave the  parameters of the systems; the  contact binaries V7 and V8 
have distances consistent with that of the cluster.
NGC~6397 appears to harbour no RR~Lyrae stars,
being its horizontal branch  remarkably blue, much like that of its analogous
cluster, M10.  To match the blue tail of the horizontal branch population, models 
of 0.64--0.66~$M_\odot$ with mass loss at the RGB are required, indicating rather thin shell masses for the HB stars.

\end{abstract}


\end{frontmatter}







\section{Introduction}
\label{sec:intro}

Galactic globular clusters, \ja{one of the most} luminous remnants of
the formation epoch of our Galaxy, play a central role in several
areas of astronomy. Being its stellar components at the same distance and of similar age, the stellar mass distribution allows the comparison of stars in different evolutionary stages,
they are tracers of the structure of the galactic halo, they
are good laboratories for the study of stellar dynamical processes,
they give key information on the early evolution of  our Galaxy
and other galaxies\ja{, and}  their ages and compositions provide severe
constraints on cosmological models. Much investigation has  
 been focused on the understanding of  the \ja{variable stars} that 
globular clusters typically harbour, in particular---but not only---the 
 RR~Lyrae stars, from the observational  and  theoretical
 points of view. \ja{These  stars  can be}
used to derive  independent estimates of several properties for individual 
stars and for the cluster as a whole.

In a series of works, mostly summarized in \cite{aaf17}, hereafter AF17, our group has
been employing CCD time series photometry of globular clusters to update 
their variable star census and to obtain independent and homogeneous estimates of their
metallicity and distance. 

 

In this paper we analyse time-series observations of \astrobj{NGC 6397} 
(C1736$-$536 in the IAU nomenclature), located at $\alpha = 17^\mathrm{h}~40^\mathrm{m}~42.^{\!\!{\mathrm{s}}}09$, 
$\delta = -53^{\circ}~40^{'}~27.^{\!\!{''}}6$ (J2000),
 $l = 338.^{\!\!\circ}17$, $b=-11.^{\!\!\circ}96$.
  With [Fe/H]~$\approx -2.0$ \ja{\citep{car09}}, it is one of the most
metal-poor globular clusters. 
Its high  King central concentration parameter ($c=2.50$, \ja{\citealt{tra95})
indicates} that it is  one of the Milky~Way clusters
that would have suffered core collapse \citep{djo86}. 
Its foreground reddening   $E(B-V)$ is  0.16--0.19~mag according to the galactic dust 
calibrations of \citet{sch11} and \citet{sch98}, and
it has an estimated age of about 13.4~Gyr \citep{gra03},
being thus one of the oldest Galactic globular clusters.
\ja{However, a recent determination by \cite{cor18}, that uses  IR~WFC3@HST
data and isochrone fittings, 
brings the cluster to
$12.6\pm0.7$~Gyr, an age akin
to that of most similar systems.} 
At $\sim2.5$~kpc \citep{gra03}, this cluster is the second  nearest to
the Sun, after  NGC~6121 (M4).

The \emph{Catalogue of Variable Stars in Globular 
Clusters}\footnote{\url{http://www.astro.utoronto.ca/\~cclement/read.html}} (CVSGC,  
\citealt{cle01}) 
lists 36  variables in the field of the cluster, of which only one or perhaps
two are of RR~Lyrae type. Variables  V1 (V639~Ara) and V2 (V825~Ara) were discovered by \cite{bai02}. 
 Star V3, an RRab variable, was mentioned by H.~H.~Swope in
    unpublished correspondence with H.~B.~Sawyer, who assigned 
    the number three in her 2nd catalogue  \citep{saw55};
    V3 is considered to be a field star and is designated as V826~Ara
    in the  \ja{\emph{Moscow General Catalogue of Variable Stars}\footnote{\url{http://www.sai.msu.su/gcvs/gcvs/vartype.htm}}
    (GCVS, \citealt{sam17}).}
    Stars V4 to 11 were announced by \cite{kal97}, although
    V10 had already been identified as the blue straggler
    BS\#11  by \cite{lau92}. Stars V10 and V11 are 
    pulsating variables of the SX~Phoenicis type. Variables
 12--24 were identified by \cite{kal03}.   Stars V15, V21,
   and V23 are SX~Phe variables; stars  V22 and V23 are
   also listed by \cite{lau92} as blue stragglers
     BS\#8 and BS\#16. Star V22, however, is listed in the
     CVSGC as a field RRc?\ variable.
     The ellipsoidal variable V16
     is the optical component of the 
     millisecond pulsar J1740$-$534; based on radial velocity
     determinations, \cite{kal03} and \cite{kal08} conclude that it is a
     cluster member. Finally, \cite{kal06}
     announced stars V25 to 36, \ja{among them one irregular variable (V26),
     two eclipsing binaries (V30 and V32), two optical counterparts
     of cataclismic variables (V33 and V34), and the rest
     variable stars of possibly assorted although unconfirmed  types.}
    
  
  \ja{A post-core-collapse, dynamically evolved globular cluster,
   NGC~6397 boasts of harbouring about two dozen  blue stragglers  near
the cluster centre, revealed by ground-based CCD photometry 
 \citep{aur90,lau92}, as well as
      at least 79
X-ray sources detected by NASA's Chandra X-ray telescope within its half-mass radius 
\citep{gri01}.} 
 Based on HST data, \cite{coh10} published optical identifications 
 for the 79  X-ray sources, among them  stars V7, V12, V13, 
 V14, V16, V17, V20, V24, V26, V30, V31, V33, V34, V35, and V36.
In particular, stars V12 and V13
     are the optical counterparts to 
     \ja{cataclismic variables} CV1 and CV6 of
   \cite{gri01}, 
 and stars V33 and 34 are the optical counterparts to CV3 and CV2,
 in which \cite{sha05} found dwarf nova-like eruptions.

  
 For the sake of completeness, it is also worth 
 mentioning \ja{these more recent developments. First,} the search for planetary transits of \cite{nas12}, who identified
12 new variables among low-main-sequence stars, all of them considered to belong to the
field, based on the proper motions and positions 
in the colour-magnitude  diagram. \ja{Second,}   the 
photometric survey by \cite{mar17} who, also among the stars of the
   lower main sequence of NGC~6397, found 412 suspected variables. \ja{And third, the   report by
   \cite{mar21} of optical modulations in the companion to the X-ray source U18, which gives support to the interpretation that it is the second ``redback'' millisecond
   pulsar known in the cluster.}

    In this paper we present a new time-series photometry
    of NGC~6397. By analysing  the light curves of
    the cluster variables, we want to obtain information on the variable stars
    and their parent system. This is the
     layout of the paper. 
    In \S~\ref{sec:observaciones}
    we present our observations and describe the DIA photometric process;
     \S~\ref{sec:CMD} deals with the colour-magnitude diagram,
    including the description of the \emph{Gaia}~DR2 method employed to 
    select the cluster members, and the matching of isochrones;
     in \S~\ref{sec:variables} we
    give the accurate identification and membership of all the
    cluster variables; \S~\ref{sec:binarias}
    is devoted to the modelling of four eclipsing binaries in the field
    of the cluster;
    in \S~\ref{sec:distance} we derive the
    distance to the cluster using several methods; in \S~\ref{sec:HB} we discuss the structure
    of the cluster horizontal branch, and present its modelling  using the Eggleton codes with mass loss during the red giant stage;  finally, in \S~\ref{sec:conclusiones}
    we give an account of the results of the work.


\section{Data and reductions}
\label{sec:observaciones}

\subsection{Observations}

The observations were performed in two sites. First,  
the Bosque Alegre Astrophysical Station
of the C\'{o}rdoba Observatory,  National University of C\'{o}rdoba, 
Argentina, whose
1.54-metre telescope 
was used  on 19 nights between June~3, 2017 and August~5, 2018.
%
%
We used a camera Alta F16M with a  detector  KAF-16803  of $4096\times4096$
square 9-micron pixels, binned $2 \times 2$, with a scale of
0.496 arcsec/pix after binning. Due to coma, the binned Bosque Alegre images were
trimmed to $1160 \times 1160$ pixels, for a useful field of
view (FoV) of  $9.7' \times9.7'$.
Second,  Las~Campanas Observatory, Chile,
whose 1.0-metre Swope Telescope 
 was used on the night of June~28, 2018, with the CCD E2V~231-84 
of $4096\times4112$ pixels also binned $2\times2$;  the scale is 0.435 arcsec/pixel after binning, and the FoV is $15.0'  \times14.9'$.
A total of 748 ($V$) and 737 ($I$)
images were acquired along the 20 nights.
The log of observations is  in Table~\ref{tab:observaciones}, where the dates,
number of frames, exposure times, and average nightly seeing are listed.

\begin{table*}
\caption{Distribution of observations of NGC~6397
\label{tab:observaciones}} 
\begin{center}
\begin{tabular}{lcccccc}
\hline 
Date & $N_V$ & $t_V$ & $N_I$ & $t_I$ & Seeing\\
 & & [s] & & [s] & [$''$]  \\
\hline
20170603     & 9  & 400 & 11 & 200 & 1.5\\
20170616     & 6  & 400 & 7 & 200 & 2.6 \\
20170630     & 8  & 400 & 9 & 200 & 1.3 \\
20170701     & 12  & 400 &12 & 200 & 1.8\\
20170728     & 12   & 400 & 12 & 200 & 1.8\\
20170818     & 9   & 400 & 11 & 200 & 2.4 \\
20170819  & 18 & 400 & 18& 200 & 1.8\\
20170820  & 24 & 90,160,300,400 & 23 & 50,60,120,200 & 1.4\\
20170826  & 5  & 120 & 6 & 60 & 1.4 \\
20170901     & 22  & 120 & 25& 60 & 1.9 \\
20170915       & 27  & 120 & 26 & 60 & 1.8 \\
20170922       & 37  & 120 & 37 & 60 & 1.3 \\
20171006       & 17  & 120 & 6 & 60 & 1.7 \\
20180518       & 63  & 40,120 & 58 & 20,60 & 1.6 \\
20180519      & 75  & 40,200 & 77 & 20,100 & 1.5 \\
20180623     & 56  & 60 & 55 & 30 & 2.2 \\
20180628$^{*}$     & 193  & 5,10 & 193 & 1,3 & 1.3 \\
20180803     & 60  & 60 & 60 & 30 & 1.3 \\
20180804      & 55  & 60 & 52 & 30 & 1.2 \\
20180805     & 40  & 60 & 39 & 30 & 1.2 \\
\hline
\end{tabular}\\
\end{center}
{\small%
 \textbf{Notes.} Columns $N_{V}$ and $N_{I}$ give the number of images taken with the $V$ and $I$
filters. Columns $\MakeLowercase{t}_{V}$ and $\MakeLowercase{t}_{I}$
provide the exposure time,
or range of exposure times. The average seeing is listed in the last column. Observations are from Bosque Alegre
Astrophysical Station, 
except the asterisked  night.}
\end{table*}

\subsection{Difference image analysis}

To carry out high-precision photometry for all  the point sources in our collection of  images of NGC~6397, 
we applied the recognised technique of difference image analysis (DIA). As in previous papers, 
we employed the  {\tt DanDIA}\footnote{\texttt{DanDIA} is built from the DanIDL library of IDL routines 
available at \url{http://www.danidl.co.uk}.} 
pipeline for the data reduction process \citep{Bramich2008, Bramich2013}.
\ja{The Bosque Alegre and Las Campanas images were reduced independently.
Reference images for the $V$ and $I$ 
filters, and for each observatory, were built by stacking the best-quality images in our collection.}  Sequences of difference images in each 
filter were then built by subtracting the respective convolved reference image from the rest of the series. 
Differential fluxes for each star detected in the reference image were measured on each difference image. 
Finally, light curves for each star were constructed
from the series of  differencial images,
by calculating the total 
flux $f_ {\mbox{\scriptsize tot}}(t)$ at each epoch $t$ from:
\begin{equation}
f_{\mbox{\scriptsize tot}}(t) = f_{\mbox{\scriptsize ref}} +
\frac{f_{\mbox{\scriptsize diff}}(t)}{p(t)} ,
\label{eqn:totflux}
\end{equation}

\noindent
where $f_{\mbox{\scriptsize ref}}$ is the reference flux,
$f_{\mbox{\scriptsize diff}}(t)$ is the differential flux, and
$p(t)$ is a photometric scale factor.
Conversion to instrumental magnitudes is achieved using:
\begin{equation}
m_{\mbox{\scriptsize ins}}(t) = 25.0 - 2.5 \log\left[ f_{\mbox{\scriptsize tot}}(t)
\right] ,
\label{eqn:mag}
\end{equation}
where $m_{\mbox{\scriptsize ins}}(t)$ is the instrumental magnitude of the star 
at time $t$. All fluxes are in ADU~s$^{-1}$, and
uncertainties are  propagated as usual. 
  The  procedure is described, with  details and caveats, by \cite{Bramich2011}.

\subsection{Photometric calibrations}

To correct for possible systematic errors, we
applied the methodology developed by \cite{Bramich2012} to solve for
the magnitude offsets to be subtracted from each photometric measurement made
on a given image.  This is a first-order correction 
to the systematic uncertainty introduced into the photometry of an image, due to an
error in the fitted value of the  scale factor $p$ (Equation~\ref{eqn:totflux}).
In the present case, the systematic error corrections were found to be negligible, always smaller than 1~mmag, except in the case of the $I$ photometry from Las~Campanas,  where a correction of 6~mmag was calculated and applied for stars brighter than the 15th magnitude.

\subsection{Transformation to the standard system}
\begin{figure*}[t] 
\includegraphics[width=\textwidth]{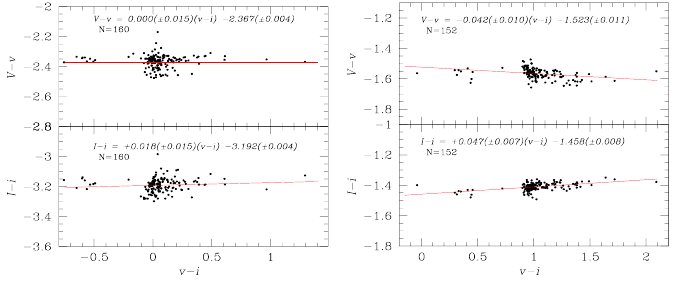}

\caption{Colour dependence of the transformations for the $V$ and $I$ filters between 
the instrumental and the standard photometric systems. The \emph{left} and \emph{right} panels correspond 
to the transformations for the observations in Bosque Alegre and Las Campanas, respectively. 
We employed a set of standard stars of \cite{Stetson2000} present in the field of NGC~6397.}
    \label{fig:transf}
\end{figure*}

Standard stars in the field of NGC~6397 are included in the online collection of Stetson (2000),\footnote{\url{http://www3.cadc-ccda.hia-iha.nrc-cnrc.gc.ca/community/STETSON/standards}} which were  used to transform our instrumental \emph{vi} magnitudes into the  Johnson-Kron-Cousins  \emph{VI} system. The mild colour dependences of the standard \emph{minus}  instrumental magnitudes are shown in Fig.~\ref{fig:transf} for  the observations from Bosque Alegre and Las~Campanas. The transformation equations are  given in the figure itself. However, given the non-significant colour dependence for the Bosque Alegre data, we opted to transform them to the standard system using the linear relations: $V = (0.980 \pm 0.004) \times v - (2.012 \pm0.082)$ and $I = (0.985 \pm 0.004) \times i - (2.926 \pm0.068)$, since they provide a good agreement with  Las~Campanas observations, of better spatial resolution.

\section{The colour-magnitude diagram}
\label{sec:CMD}

\subsection{Cluster stellar membership}
\label{membership}

\begin{figure*}  
\centering
\includegraphics[width=0.47\textwidth]{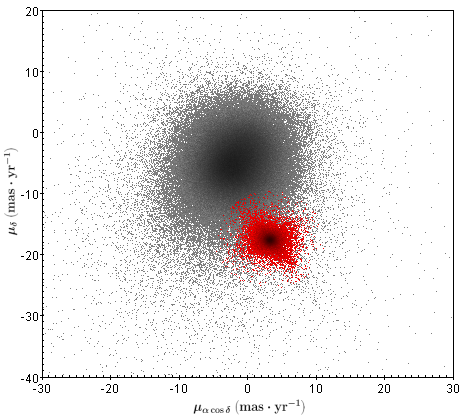}
\includegraphics[width=0.5\textwidth]{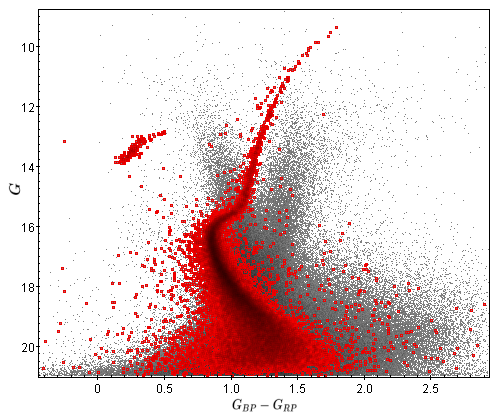}
\caption{Membership of stars in the NGC~6397 field, based on \emph{Gaia}~DR2 data
(see \S~\ref{membership}).
The \emph{left} panel shows the Vector-Point Diagram (VPD) for 472,304 stars within 
60 arcmin around the cluster. Red dots represent stars considered likely  members. 
The cluster clearly stands out against the vast number of field stars. 
The \emph{right} panel displays the  colour-magnitude diagram in the \emph{Gaia} photometric  system.}
    \label{fig:VPD}
\end{figure*}

We applied the
 method of \cite{bf19} to identify
probable members in the field of the cluster. It uses the high quality astrometric data available
in \emph{Gaia}~DR2, implementing the Balanced Iterative
Reducing and Clustering using Hierarchies
(BIRCH) algorithm \citep{Zhang1996}  in a four-dimensional
space of physical parameters---positions
and proper motions---that detects groups of stars in
the 4D-space. In this way we extracted 31,047  \emph{Gaia} stellar sources that
are  likely cluster members, from  a much larger  sample of 472,304 stars within 60 arcmin around 
the cluster. A total of 17,482  stars are in the field of our Las~Campanas images,
of which 12,971 are considered members.
Figure~\ref{fig:VPD} shows
the Vector-Point Diagram (VPD) of cluster and field stars,  and the corresponding 
\emph{Gaia} colour-magnitude diagram.

\subsection{Cluster reddening}
\label{reddening}

NGC 6397 belongs to  a group of globular clusters located toward the centre of the Milky Way; hence, it is affected 
by  interstellar reddening of a patchy nature. To  differentially deredden our photometry, we employed the differential 
reddening grid calculated by \citet{AlonsoGar2012}, that covers nearly the whole field of our images from the 
Swope telescope. \ja{The resolution of the grid is 5.4$''$ ($0.^{\!\!{\mathrm{s}}}36$) in RA and 3.6$''$
in DEC.} 
The procedure to deredden each star starts by first averaging the four 
neighbouring differential reddening values given in the grid, and then adding a  mean overall reddening. The mean value of $E(B-V)$  suggested by \citet{AlonsoGar2012} for NGC~6397 is $0.18$. The Galactic 
dust distribution and the calibrations of  \cite{sch11} and \cite{sch98}  suggest values of 0.166 and  $0.188\pm0.003$, respectively.
In Fig.~\ref{fig:CMD} we show the CMD before (left panel) and after (right panel) the reddening correction and membership cleaning.
In the latter we adopted $E(B-V)=0.19$ and a distance of 2.5~kpc, to place the two 
isochrones for 13.0 and 13.5~Gyr. To deredden the colour $(V-I)$ we employed the ratio $E(V-I)/E(B-V) = 1.259$. Although differential reddening is mild across the NGC~6397 area, a sharper distribution 
of stars is noticeable after this correction.



\subsection{Isochrones and ZAHB fitting}
\label{sec:isocronas}

\begin{figure*}
\centering
\includegraphics[width=\textwidth]{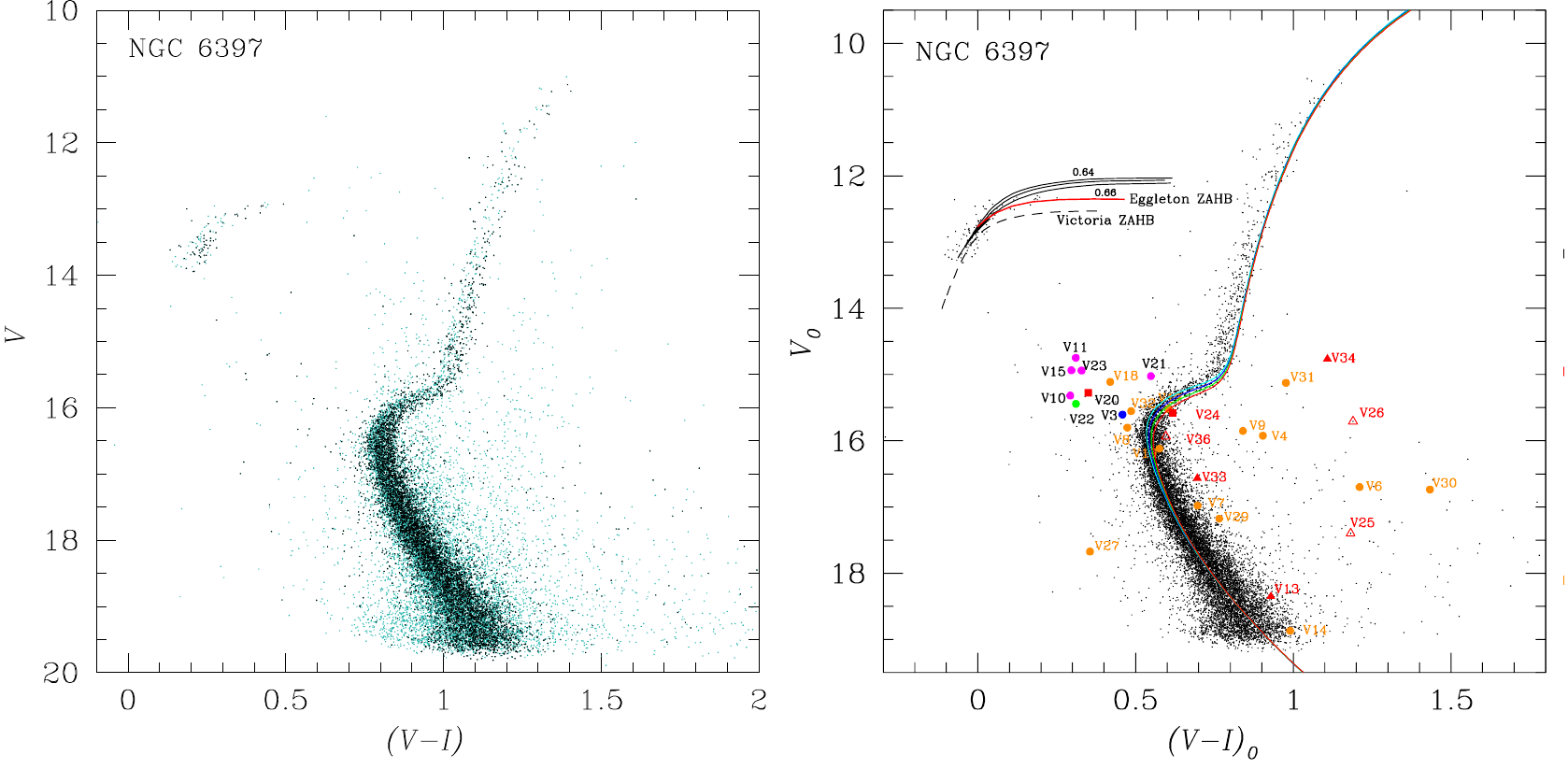}
\caption{Colour-magnitude diagram  of NGC~6397, built from Las~Campanas data. The \emph{left} panel shows 
the observed, reddened stellar distribution. Black and blue dots represent cluster members and field stars 
respectively, according to the membership analysis of $\S$~\ref{membership}.  In the \emph{right} panel  only  cluster members 
are plotted, \ja{after their colours were individually corrected by differential reddening 
 ($\S$~\ref{reddening}),  and  globally shifted  an average colour excess  $E(B-V)=0.19$.
The star sequences appear well matched by four
isochrones  calculated using the models of \citet{Vandenberg2014},  for ages (left to right): 12.0, 12.5, 13.0, 
 and 13.5~Gyr, and  [Fe/H]~$= -2.0$, 
  $Y = 0.25$, and [$\alpha$/Fe]~$ = +0.4$. The isochrones were 
 located at a distance of 2.5~kpc.} 
At the HB, three continuous black loci, \ja{also positioned at 2.5~kpc,} show the evolutionary tracks for total masses of 0.64, 0.65, and  0.66~$M_{\odot}$, 
calculated as explained in  \S~\ref{sec:HB}. The resulting ZAHB is shown in red, and for comparison we include, 
as a dashed line, the ZAHB computed from \citet{Vandenberg2014} models, for the same  parameters.
The known variables  measured in the present study are labelled   with different symbols and colours: 
RRab (\emph{blue}); RRc (\emph{green}); SX~Phe (\emph{lilac}); $\gamma$~Dor-type?\ and cataclysmic (\emph{red}); contact binaries and
ellipsoidal (\emph{orange}).}
    \label{fig:CMD}
\end{figure*}

The colour-magnitude diagram (CMD) of NGC~6397, built with data from the Swope telescope (the
better quality data) is shown in Fig.~\ref{fig:CMD}.  In the left panel the member and 
non-member stars are represented by black and light blue points. In the right panel we present 
the  CMD after the non-members have been removed, and the member stars have been subjected to the differential 
reddening correction ($\S$~\ref{reddening}). The CMD diagram appears very clean, which enabled 
a proper matching of  isochrones and ZAHB models. These best fitting models were taken from the grid of the Victoria-Regina models of 
\citet{Vandenberg2014} for  [Fe/H]~$= -2.0$, $Y = 0.25$, and [$\alpha$/Fe]~$ = +0.4$, and   
ages of 13.0 and 13.5~Gyr. The isochrones  match very well the stellar unreddened distribution for 
a distance of 2.5~kpc and $E(B-V)=0.19$.
It is pertinent to note here that  \cite{br18} recently derived, from  the $V$ luminosity at the main-sequence turnoff, 
an absolute  age of $13.4 \pm1.2$~Gyr for the cluster.

In the CMD it is evident a very blue horizontal branch  void of any RR Lyrae stars. 
The only two RR Lyrae known in the field of the cluster (V3 and V22) are clearly non-members
and background objects, as they  appear in the blue straggler area, near the turnoff. The structure of the HB will be discussed in $\S$~\ref{sec:HB}.

\section{Variable stars in the field of NGC 6397}
\label{sec:variables}

\begin{figure*}  
\centering
\includegraphics[width=0.49\textwidth]{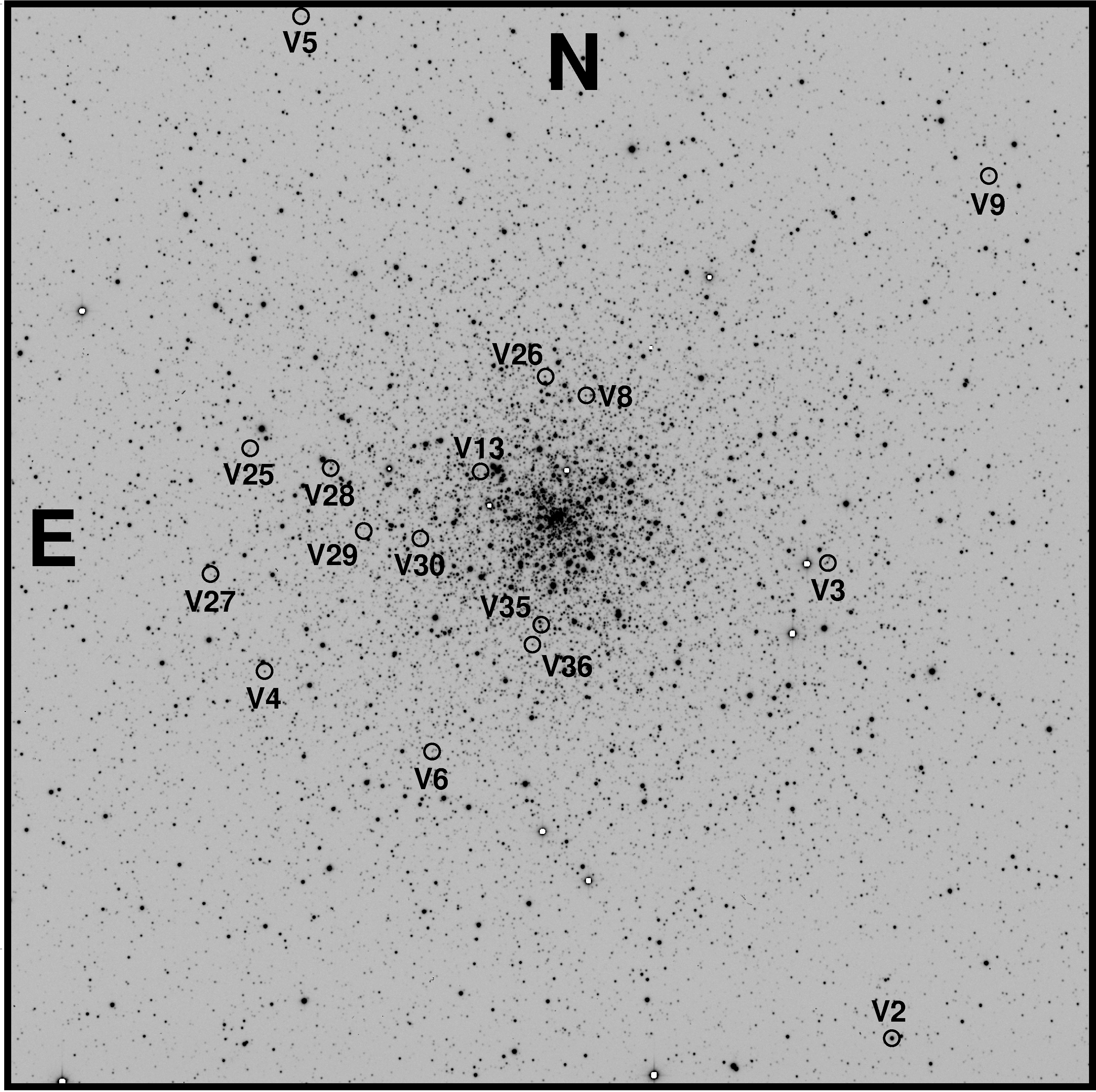}
\includegraphics[width=0.49\textwidth]{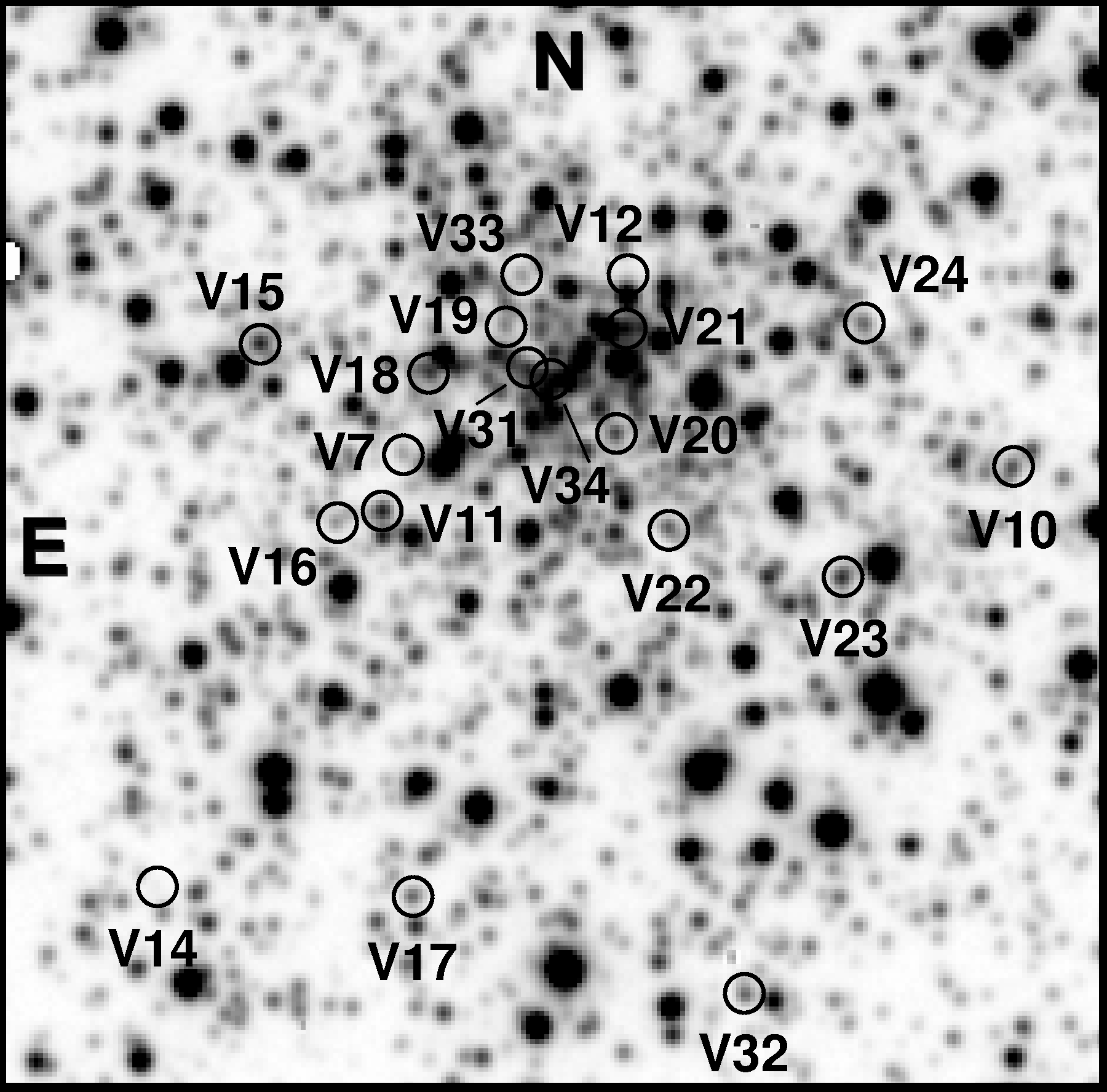}
\caption{Identification charts for the variables in NGC~6397. \emph{Left}, the periphery of the cluster  ($15'\times15'$). \emph{Right}, the cluster core ($1.7'\times1.7'$). Only star V1 is outside the larger field.}
    \label{fig:cartas}
\end{figure*}

The 36 stars catalogued in the CVSGC \citep{cle01}, their coordinates and \emph{Gaia}~DR2 identifications 
     (when available), and their membership as estimated \ja{in this paper,} 
     are listed in Table~\ref{tab:identificaciones}. Among them, only V1 and V3 appear flagged as variables in
     \emph{Gaia}~DR2. Finding charts for all   variables, except V1, are in Fig.~\ref{fig:cartas}.
     
\begin{table*}

	\caption{Variable stars in the field of NGC~6397 \label{tab:identificaciones}}
	\begin{center}
	\begin{tabular}{llccccc} 
		\hline
  Variable Id. & Variable Type$^{a}$ & $V$$^{a}$ &RA$^{b}$ & Dec$^{b}$ &\emph{Gaia}~DR2 Source Id. & Memb.$^{c}$ \\\
      &   & [mag] & (J2000.0) & (J2000.0)& \\
\hline
V1 & M & 13.36 & 17:41:04.75 &$-53$:32:58.6 & 5921753573179938304 &m? \\ 
V2 & RV? & 12.88  &17:40:10.86&$-53$:47:33.9 & 5921745812182394496  &f\\   
V3 & RRab & 16.07  & 17:40:16.98 & $-53$:41:03.6 & 5921748045565559040  &f \\
V4 & EW & 16.35 & 17:41:08.83 &$-53$:42:34.3 &  5921750072784246400  &f\\
V5 & E&18.70 &17:41:05.55 &$-53$:33:36.0 &  5921753573174589440  & \ja{f}   \\ 
V6 & EW & 17.16 &  17:40:53.35 & $-53$:43:39.6  &  5921744884469398656 &f  \\ 
V7 & EW & 17.09 & 17:40:43.93 &$-53$:40:35.5 &5921745296766522752 & m  \\
V8 & EW & 16.24 & 17:40:39.27 &$-53$:38:47.2 & 5921754161593008384 & m \\
V9 & EW & 16.25& 17:40:02.25 & $-53$:35:45.3 & 5921755295470448256 & f \\
V10 & SX Phe & 16.00  &17:40:37.55 & $-53$:40:36.5 & 5921748217344441856  &m \\
V11 & SX Phe & 15.43 & 17:40:44.14 & $-53$:40:40.8 & 5921745296766500224  &m \\
V12$^{d}$ & CV? & 17.52 & 17:40:41.62 & $-53$:40:20.0 & $\dots$ & $\dots$ \\ 
V13 & CV? & 19.43 & 17:40:48.94 &$-$53:39:49.5 & 5921751137936013056  & $\dots$ \\   
V14 & E & 19.25 & 17:40:46.50 &$-$53:41:15.7 & $\dots$ & $\dots$ \\  
V15 & SX Phe & 15.43 & 17:40:45.42 & $-53$:40:25.3 & 5921751172297381120 & m\\
V16 & ELL & 16.65  & 17:40:44.63 &$-53$:40:41.9 & 5921745296766500864  & m\\
V17 &ELL? & 16.23 & 17:40:43.82 &$-53$:41:16.6 &5921745296766429824 & m\\  
V18 & E & 15.75 & 17:40:43.64 &$-53$:40:28.0 & 5921751172301644032  &$\dots$\\
V19 &E &17.12 & 17:40:42.85 &$-53$:40:23.8 & 5921751172301639808   &$\dots$\\
V20 & $\gamma$ Dor? &15.83 & 17:40:41.70 & $-53$:40:33.6 &5921745296786376320 & m\\
V21 & SX Phe &15.47 & 17:40:41.59 &$-53$:40:23.9 & 5921745296786384128 & $\dots$\\
V22 & \ja{RRd$^c$} & 16.15 & 17:40:41.15 &$-53$:40:42.3 & 5921745296766606336  & \ja{f} \\
V23 & SX Phe& 15.52 & 17:40:39.34 &$ -53$:40:47.0 & 5921748251704126464  & $\dots$\\
V24$^e$ & $\gamma$ Dor? &16.17 & 17:40:39.10& $-53$:40:23.4 & $\dots$ & $\dots$\\
V25 & ? & 17.98 & 17:41:10.18 &$-53$:39:30.9 &5921750652596773888 &f \\
V26 & ? &16.24 &17:40:43.05 &$-53$:38:31.7 & 5921751305431697920 &f\\
V27 &ELL? & 18.25 & 17:41:13.81& $-53$:41:14.5& 5921750244578290816 &f\\
V28 & ? & 15.13 & 17:41:02.73 &$-53$:39:47.5 &5921751000496772224 &f\\
V29 &ELL? &19.72 & 17:40:59.67 &$-53$:40:39.0 &5921751034856719488 &$\dots$\\
V30 &E&17.64 &17:40:54.53 & $-53$:40:44.7 & 5921750931784538112 &f\\
V31 &ELL? &15.99 & 17:40:42.62 &$-53$:40:27.5 & 5921745296766546176 &$\dots$\\
V32 &E&16.13 & 17:40:40.33 &$-53$:41:25.6&5921745262406760960 &m\\ 
V33 &CV? & 18.3 & 17:40:42.71& $-53$:40:18.7 & 5921751172296037632  &$\dots$\\
V34 & CV &16.2 &17:40:42.40 & $-53$:40:28.7 & 5921745296766570368 &$\dots$\\
V35$^{f}$ & ELL &18.79 & 17:40:43.35 &$-53$:41:55.6&$\dots$&$\dots$\\  
V36 & ? &16.53 &17:40:44.14 &$-53$:42:11.7 & 5921745155038332032& m\\
		\hline
	\end{tabular}\\
	\end{center}
	{\small
	\textbf{Notes.}\\
	$^a$CVSGC \citep{cle01}.\\
	$^{b}$Coordinates are mostly from \emph{Gaia}~DR2, and the rest are from the  HST archive image \texttt{u5dr0401r}. \\
	$^{c}$\ja{This work.} \\
	$^{d}$Bright \emph{Gaia} source 5921745296786384896 is at $1''$ towards the South.\\
	$^{e}$\emph{Gaia} source 5921748251704160512 is at $0.7''$ towards the NE.\\
	$^{f}$Bright \emph{Gaia} source  5921745262406612608	is at  $1.3''$ towards the SE.}  
\end{table*}

Several of these stars, such as 
V12, 13, 14, 33--36, are in the observed field, but  are too faint
or too near a bright companion  \ja{to be properly detected or measured in our data.
Besides, stars V2, 5, 6, and 9 are outside the trimmed Bosque Alegre 
field, i.e., they were observed during only one night
at Las~Campanas (cf.\ Table~\ref{tab:observaciones});
in that night, V2 and V6 did not show any variations.}
We shall therefore limit our light-curve  analysis  to
those stars---listed in Table~\ref{tab:variables}---of variability clearly detected with our photometry.
 \ja{In all cases we attempted a new determination of the period based on our data, using 
the string-length method \citep{dwo83}.
However, a trial and error approach showed that, in some 
cases, the light curve was better phased with the period given in the CVSGC.}
In Table~\ref{tab:timeseries} we report the $V$ and $I$ time-series photometry
for all variables in our field of view; the full table,
which also includes the parameters of Equation~\ref{eqn:totflux}  computed by
{\tt DanDIA},
is  published only in
electronic format.
Individual variables are discussed
next, and the light curves are shown in Figs.~\ref{fig:LCs} and \ref{fig:SXa}.

\begin{table*}
	\caption{Data on the variable stars of NGC~6397 reported in this work\label{tab:variables}}
	\begin{center}
	\begin{tabular}{ll c c c c c c} 
		\hline
  Variable  Id.  & Variable Type
 &  $<V>^{a}$ & $<I>^{a}$ & $A_V$ & $A_I$ & $P$ & HJD$_\mathrm{max}$ \\
       &  &[mag] & [mag] & [mag] & [mag] & [d] & $+$245 0000.  \\
\hline
V3 & RRab & 16.176  &15.526&1.222&0.895&0.494748&8257.9189 \\
V4 & EW & 16.537 &15.390&0.667&0.585&0.422742&8258.8602 \\
V5 & E&18.853 &16.167&1.057&0.681& \ja{0.27$^c$} &8297.7744 \\ 
V7 & EW & 17.657 &16.694&0.905&0.836& \ja{0.2699$^c$} &8257.8506  \\
V8 & EW & 16.414 &15.713&0.421 &0.431&0.271240&7963.7078 \\
V10 & SX Phe & 15.907$^{b}$ &15.375$^{b}$&0.070&$\dots$&0.030030&7908.8962  \\
V11 & SX Phe & 15.340 &14.794&0.053&0.035&0.038262&8033.5452 \\
V15 & SX Phe & 15.525$^{b}$ &14.990$^{b}$&0.070&$\dots$&0.025240&8258.8130 \\
V16 & ELL & 16.767 &15.741&0.123&0.149&1.489390&8257.8884 \\
V18 & E & 15.701$^{b}$ &15.042$^{b}$&0.218&0.167& \ja{0.7867$^c$} &8297.6893 \\
V20 & $\gamma$ Dor? &15.863$^{b}$&15.275$^{b}$&$\dots$&$\dots$& \ja{0.8612$^c$}  &8033.5452 \\
V21 & SX Phe &15.634 &14.828&0.363&0.114&0.038961&7984.6207 \\
V22 & \ja{RRd} & 16.059 &15.472&0.184&0.123& \ja{0.46497} &8258.7534  \\
 &&&&&& \ja{0.30669} &\\
V23 & SX Phe& 15.523 &14.961$^{b}$&0.081&$\dots$&0.038005&8033.5091  \\
V24 & $\gamma$ Dor? &16.174$^{b}$ &15.316$^{b}$&0.123&$\dots$& \ja{0.4572$^c$} &8033.5398 \\
V25 & ? & 17.987$^{b}$ &16.566$^{b}$&0.190&$\dots$& \ja{1.2306$^c$} &7998.5528 \\
		\hline
	\end{tabular}\\
\end{center}
{\small
\textbf{Notes.}\\
	$^a$Intensity-weighted mean.\\
	$^b$Magnitude-weighted mean ($\overline{V}$, $\overline{I}$)}. \\
		\ja{$^c$CVSGC \citep{cle01}.}\\
\end{table*}

\begin{table*}[t]
	\caption{Time-series $V$ and $I$ photometry for the variables in our
field of view 
\label{tab:timeseries}}
\begin{center}
\begin{tabular}{lccccc} 
		\hline
  Variable Id.& Filter & HJD & $m_\mathrm{std}$ & $m_\mathrm{ins}$ & $\sigma_m$ \\ 
  & & [d] & [mag] & [mag] & [mag] \\ 
		\hline
		V2 & $V$ & 2458297.56732 & 13.249 & 14.943 & 0.004 \\ 
V2 & $V$ &2458297.57256 &13.242& 14.937& 0.005 \\ 
	\vdots&\vdots&\vdots&\vdots&\vdots&\vdots \\ 
	V2 & $I$ &2458297.58010 & 9.556& 10.820& 0.002 \\ 
 V2&$I$& 2458297.58886&  9.552& 10.816& 0.002 \\ 
	\vdots & \vdots &\vdots&\vdots&\vdots&\vdots \\ 
		V3 & $V$ &2457908.85026& 16.365& 18.750& 0.006 \\ 
 V3&$V$&2457908.85498& 16.377& 18.762& 0.012 \\ 
	\vdots&\vdots &\vdots&\vdots&\vdots&\vdots \\ 
	V3 & $I$ &2457908.84190& 15.620& 18.825& 0.012 \\ 
V3&$I$&2457908.84429& 15.626& 18.831& 0.012 \\ 
	\vdots&\vdots &\vdots&\vdots&\vdots&\vdots\\ 
		\hline
	\end{tabular}\\
\end{center}
{\small
\textbf{Notes.} The columns  $m_\mathrm{std}$ and $m_\mathrm{ins}$ are the standard and instrumental
magnitudes,  and $\sigma_m$ is the error (the same) for both.
This is an excerpt from the full
table, which is available only with the electronic version of the article.}
\end{table*}


\subsection{RR Lyrae stars}

\textbf{V3.} This is the first RRab star known in the NGC~6397 area; however, there is firm evidence  
that it does not belong to the cluster. This is confirmed by its position in the CMD, near the
cluster turnoff.

\textbf{V22.} The second RR~Lyrae variable present in the cluster area, it also appears in
the blue straggler zone of the CMD  and was originally identified as one of them by \cite{lau92}. 
In the CVSGC the star is listed as a probable double-mode RR Lyrae, but no periods are suggested. 
\ja{While \cite{kal03} discussed the possibility of V22 being a pulsating
multiperiodic variable related to $\gamma$~Doradus stars,  \cite{kal06} classified it as an RRd, with $P_0=0.52$~d
and $P_1 = 0.344$~d ($P_1/P_0=0.66$).
In our analysis we found that a period $P_0=0.46497$~d produces the  light curve 
shown in Fig.~\ref{fig:LCs}; there is, however, evidence of a second period $P_1$ of 0.30669~d, with which we also derive
 a ratio $P_1/P_0=0.66$. Both periods are listed in Table~\ref{tab:variables}.} 

\subsection{SX Phoenicis stars}

\textbf{V10.} With V11, this star was originally identified
as an SX~Phe variable by \cite{kal97}. A low amplitude modulation with a periodicity of about 0.287~d is evident in the data from Las~Campanas. On top of this, clear short-period variations with a period of 0.030~d are detected. This period agrees with that reported in the CVSGC (0.0308~d). An amplitude modulation of the short period variations is explained by the presence of a secondary frequency of period 0.03378~d, which is likely a non-radial mode. A model built with the said three periods is shown in Fig.~\ref{fig:SXa}. 

\textbf{V11.} Clear monoperiodic variations are observed in the light curves (Figs.~\ref{fig:LCs} and \ref{fig:SXa}), phased with a period of 0.038262~d \citep{kal06}. 

\textbf{V15.} Identified as an SX~Phe variable by \cite{kal03},  with V21 and 23; all of them  are in
the blue straggler region of the CMD. The small variations in brightness  
are displayed in Fig.~\ref{fig:SXa}. The two periods 0.02524~d and 0.02200~d represent rather well the 
data periodicity and the amplitude modulations. This
is a very short period for an SX~Phe variable. \cite{kal03} found for V15
a period of 0.0215~d and discussed the possibility that it is 
a pulsating hot dwarf rather than an SX~Phe.

\textbf{V21.} Mild variations in the $V$-band are seen in this SX~Phe star which are properly folded  with a period $P=0.038961$~d (see Fig.~\ref{fig:LCs}). This period represents well the periodic behaviour of the variations, but some mean magnitude modulations are also seen in Fig.~\ref{fig:SXa}, which may be due to a secondary periodicity not detected in our data.

\textbf{V23.} Small-amplitude variations are clear in this SX~Phe star, showing a periodicity of 0.03805~d
(Fig.~\ref{fig:SXa}).  

\subsection{$\gamma$ Doradus stars?}

In the CMD (Fig.~\ref{fig:CMD}), V20 and V24 are found among the blue straggler stars and in the main sequence near the turnoff point, respectively. Light variations of these two stars were reported by \citet{kal06}, giving periods of 0.8612 and 0.4572~d. These rather long periods for a typical SX~Phe star led these authors to speculate  that V20 and 24 belong to a new type of Population~II variables, counterpart of the Population~I  $\gamma$~Doradus class, i.e., nonradially oscillating stars  locate 
near the intersection of the red edge of the classical instability strip and
the main sequence.

\textbf{V20.} 
Due to seeing conditions and faintness of the star, it could not be measured in  Bosque Alegre images, although it was observed  at Las~Campanas---albeit only five hours. The Swope light curve is in Fig.~\ref{fig:LCs}. A mild variation consistent with the period 0.8612~d is clearly seen, 
\ja{consistent with} the variability and period of \citet{kal06}.

\textbf{V24.} Like V23, it presents small-amplitude variations of the SX~Phe type.
Our data are not properly phased with the period 0.4572~d reported by \cite{kal06},
but the combination of this value and the 
SX~Phe-like, short-period of 0.02754~d produce the rather good representation shown  
in Figs.~\ref{fig:LCs} and \ref{fig:SXa}.
In the cluster CMD (Fig.~\ref{fig:CMD}), the star is  towards the red edge of the main sequence which, together with its proper motion, make clear that V24 does not belong to the cluster. 
There is a star of similar brightness at $0.7''$ towards the NE, \emph{Gaia}
source 5921748251704160512 (cf.~Fig.~1 of \citealt{kal03}). \ja{Star V24, on the other hand, is not a \emph{Gaia} DR2 source.}

\subsection{Eclipsing Binaries and Ellipsoidal Variables}

\textbf{V4, V7,  V8.} Neat eclipses are observed for these three contact binaries or EW-type stars; the periods are 0.42274~d, 0.2699~d, and 0.27124~d, respectively. Star V7, discovered and identified as an W~UMa binary by \cite{kal97}, is not detected in our Bosque Alegre observations. Our (Las~Campanas) period for V7 coincides with Kaluzny \& Thompson's  \citeyearpar{kal03}. In \S~\ref{sec:binarias} we present and discuss a modelling  of these light curves.

\textbf{V5.} Its Swope light curve in $I$ shows clear eclipses; the one in $V$, however, is much noisier (Fig.~\ref{fig:LCs}). 
They were phased with the period given in the CVSGC,  $P=0.27$~d. The  star is too red to appear in our CMD. 
An analysis  of the light curves of this binary is  presented in \S~\ref{sec:binarias}.

\textbf{V6, V9,  V14.} \ja{No variations were detected in our one-night data from Las~Campanas  for V6 and V14. On the other hand, the
Swope light curve of the eclipsing binary V9  is fairly incomplete (Fig.~\ref{fig:LCs}) and therefore was 
not analysed. Nevertheless, the three stars are  included in the CMD.}
 

\textbf{V16.} This ellipsoidal variable is the optical component of the 
     millisecond pulsar J1740$-$5340---see discussion in \cite{kal03} and \cite{kal08}. In Fig.~\ref{fig:LCs}
 our light curve has been folded with the period listed by \cite{kal06}, i.e.,  1.35406~d.
 
\textbf{V17.} This star is listed as a probable elliptical variable in the CVSGC.
In our data, however, we did not detect any variations. It appears near the cluster turnoff.

\textbf{V18.} This eclipsing variable is a cluster blue straggler. \cite{kal03} reported an
unusual light curve (see their Fig.~10); they propose that it is a detached or semidetached system 
composed of stars with very similar surface brightness, rather than a contact binary. \ja{Our light curves
have been phased with $P=0.78669$ \citep{kal06}.} 


\textbf{V19.} 
No variations are noticed in our data. We therefore   cannot confirm its reported variable nature,
which appears to be of rather small amplitude (cf.~Fig.~8 of \citealt{kal03}).


\subsection{Cataclismic Variables}

\textbf{V12, V13, V33,  V34.} These stars were identified by \cite{kal06} with the  cataclismic variables CV1, CV6, CV2, and CV3 of \cite{gri01}, respectively. Although they are  rather faint  for the precision of our photometry, or  are blended with a brighter neighbour 
(V12), 
we were nevertheless   able to detect and measure  stars V13, V33, and V34, and include them in the CMD. However, we could not find any variations. The mild, long term or sporadic variations given by \cite{kal06}---their Fig.~3---rather explain the lack of variations in our data.

\subsection{Others}
\textbf{V2.} According to the CVSGC, this is a long-period 
(39.5977~d) variable of possibly RV~Tau type. We covered it
during only one night at Las~Campanas; no variations were observed in its light curves, 
from which we estimated  $<V>=13.25$ and $<I>=9.55$. Its colour is too red to be represented  
in our CMD.

\begin{figure*} 
\includegraphics[width=\textwidth]{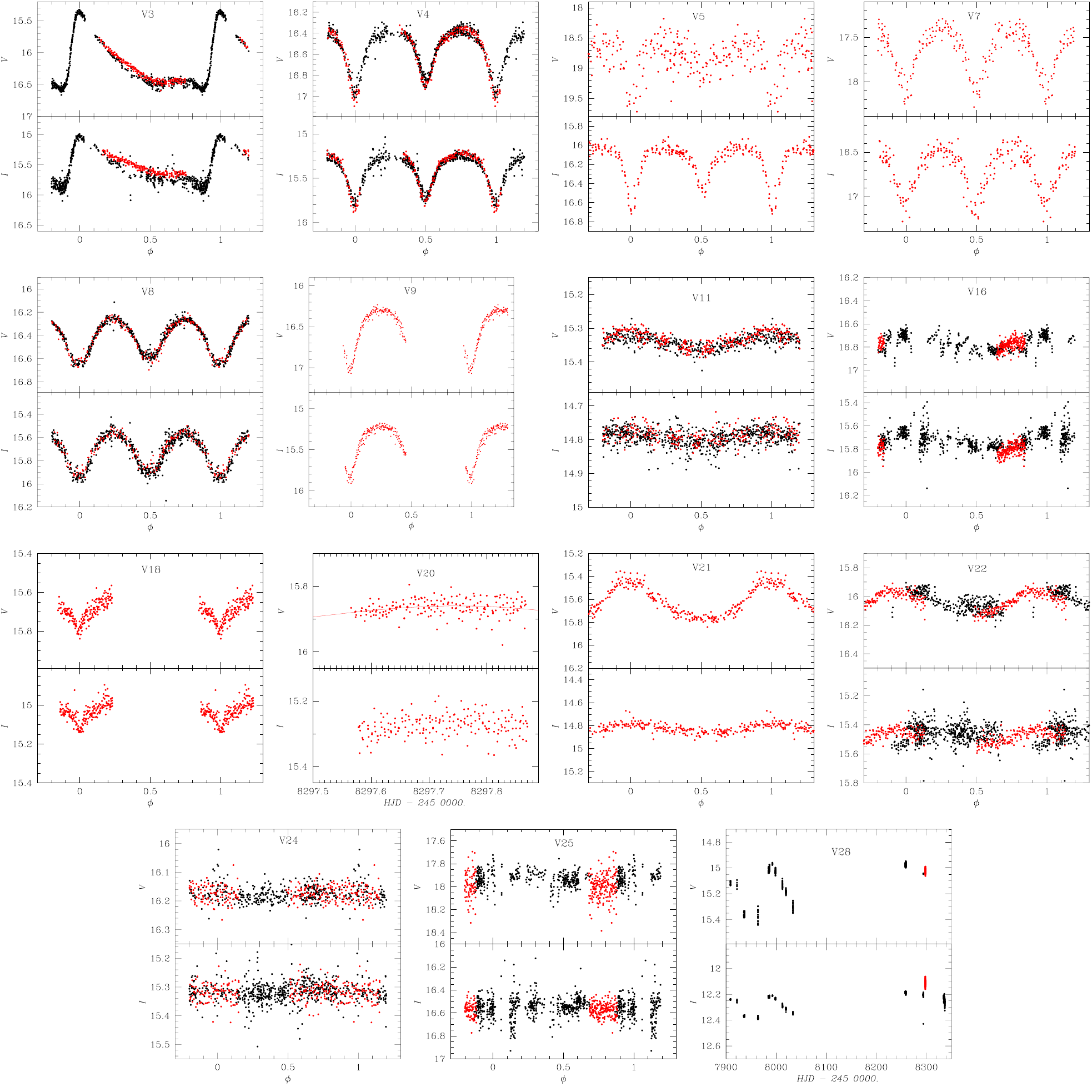}
\caption{Light curves in \emph{VI} of some variables in the field of NGC~6397. Black and red symbols stand for Bosque Alegre 
and Las~Campanas observations, respectively. \ja{Note that V20 and V28 are not phased but plotted vs.\ HJD; 
the solid curve in V20 is the portion of a $P = 0.8612$~d model.}}
    \label{fig:LCs}
\end{figure*}

\begin{figure*} 
\centering
\includegraphics[width=\textwidth]{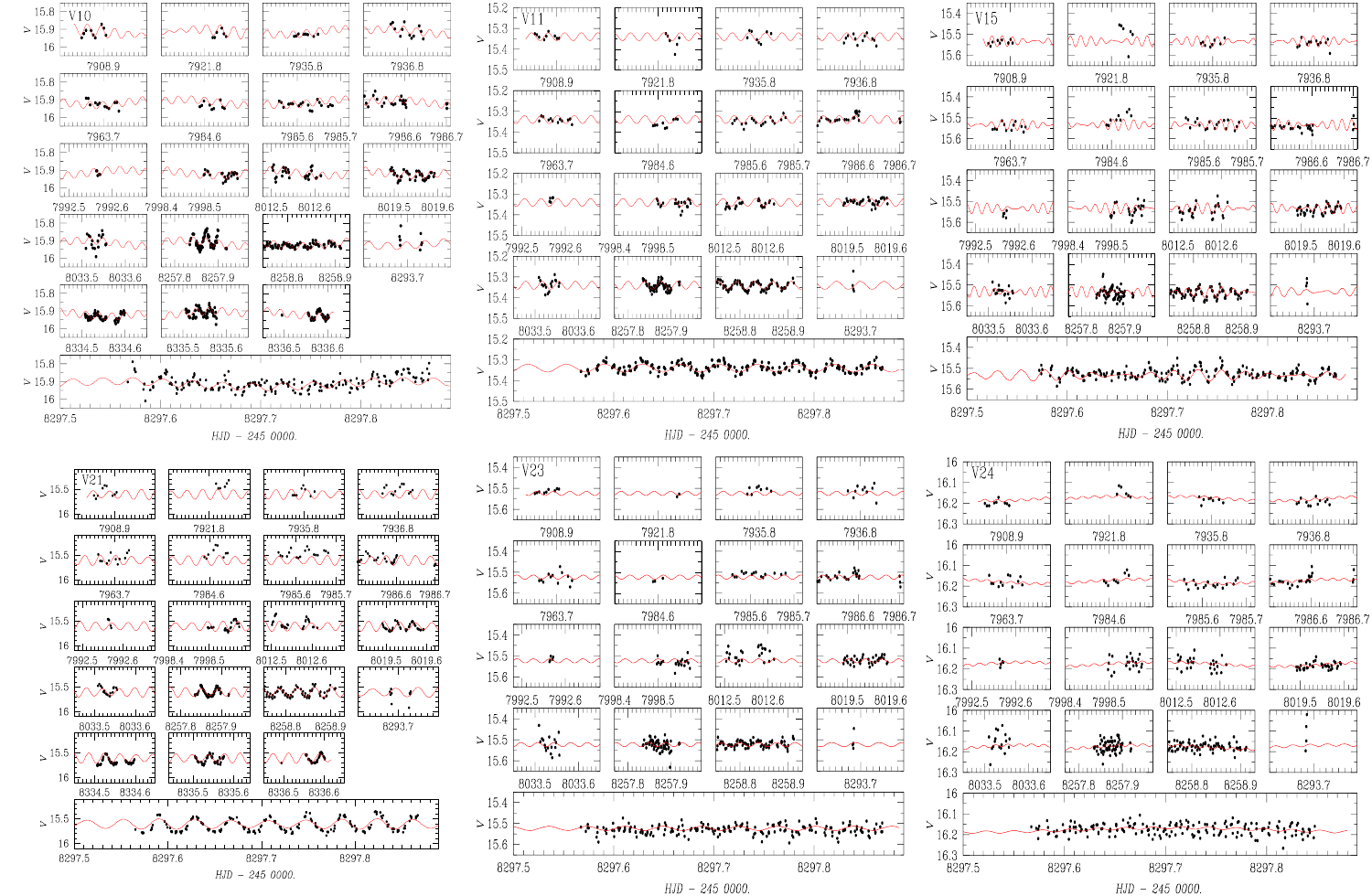}


\caption{Light curves in $V$ of six SX~Phoenicis stars in the field of  NGC~6397. \ja{The long boxes contain $\sim7.5$ hours of Swope data.}}
    \label{fig:SXa}
\end{figure*}

\textbf{V25.} The light curve (Fig.~\ref{fig:LCs}) was phased with a period of 1.2306~d \citep{kal06}. 
The slight variations are not conclusive as to   how  classify  this star.

\ja{\textbf{V28.} We could not phase properly the light curve of this rather bright 
($V\approx 15.2$~mag) star with the period  25.997~d suggested by \cite{kal06}.
A plot of $V$ vs.\ HJD (Fig.~\ref{fig:LCs}) shows clear variations, and one full cycle suggests 
a much longer period of  $P=97.78$~days.
This new period is not rigid but is rather a characteristic time of variation, as
stochastic variations are hinted by our data. 
The star is very red and is outside the CMD limits of Fig.~\ref{fig:CMD}. 
It was classified by \cite{kal06} as a pulsating star, but  could in fact be a spotted variable. 
}

\subsection{Variables not detected}

\ja{\textbf{V26, 27, 29--32, 35--36.} We were not able to detect variations using published periods or after searching for
them with the string-length method. They are nevertheless represented in our CMD, with the exception of V35, which 
is too faint and too near a bright star.}

\section{Modelling contact binaries in  NGC~6397}
\label{sec:binarias}

Figure~\ref{fig:LCs}   shows  the $V$, $I$ light curves of the eclipsing binaries
V4, V5, V7, and V8, among other stars.   They evidently have  morphologies typical of semidetached and contact
binaries. To model the systems, we used the code \textsc{BinaRoche} \citep{laz09},  which 
 simultaneously fits both the $V$ and $I$ light curves. 

The \textsc{BinaRoche} model takes into account the deformation of the stars, in the Roche model
 geometry, and the inhomogeneous temperature distribution over the stellar surfaces, considering
  gravity darkening and heating effects. 
 The gravity darkening exponents can be declared free parameters, but
 the standard values---$\beta = 0.25$ for a star with a radiative envelope, 
and $\beta = 0.08$ if the star has a convective envelope---usually produce
satisfactory fits. The codes adopts $T \propto g^\beta$, with $T$ and $g$ being the local effective temperature
 and surface gravity, respectively.
The filling factor $S_i$ of each star, which is also a free parameter of the model, is
defined as the ratio between its polar radius ($r_p$) and the polar radius of its critical
 Roche lobe ($R_p$): $S_i= (r_p/R_p)_i$, $i= 1, 2$. 
The mass of the primary component---in solar mass units---and  the mass ratio $q= M_2/M_1$ are
parameters of the model, since the stellar surface fluxes depend on
$T_{\mathrm{eff}}$, $\log g$, and [Fe/H]. The code adopts stellar atmospheres fluxes from
the BaSel library \citep{lej98}.
 Expressing the masses of the binary components in 
solar masses allows the derivation of the distance to the system in parsecs. 

The analysis starts assuming
the binaries are members of NGC~6397, i.e, they have  the cluster's metallicity and
intestellar extinction. 
%
\ja{The adoption of a $E(B-V)$ value, and the relative contribution of the primary
component to the total flux in $V$ and $I$, a result of the light curves fits, allows
us to obtain the intrinsic colour $(V-I)_0$ of the primary star from the observed
colour of the binary at the maximum. Then, the effective temperature $T_{\mathrm{eff},1}$ of
the primary is estimated from the calibrations $(V-I)_0$ vs.\  $T_{\mathrm{eff}}$ of \cite{hua15}  
if $-0.8 \leq \mathrm{[Fe/H]} \leq +0.3$, or \cite{cas10}   when [Fe/H]= -2.0.}

To estimate the mass of the primary component, we  adopted the
 relation of \cite{tor10} for stars in detached,
 non-interacting binary systems with accurately measured masses and radii. 
 They give polinomial fits for $\log M (M_\odot)$
 and $\log R (R_\odot)$, in terms of $T_{\mathrm{eff}}$, $\log g$, and [Fe/H]. Unfortunately,
  these relations do not include systems with 
 metallicity as low as $-2.0$~dex, i.e., that adopted for NGC~6397, which  introduces an additional
 uncertainty. 
  The mass of the secondary star is derived from the free parameter $q= M_2/M_1$.
Since the relative flux of the primary and $\log g_1$ are results of the model,
a few iterations are necessary to reach  stable values of $M_1$ and
$T_{\mathrm{eff,1}}$.

Next we discuss the individual systems.

\begin{figure}
\centering

\includegraphics[width=0.42\textwidth]{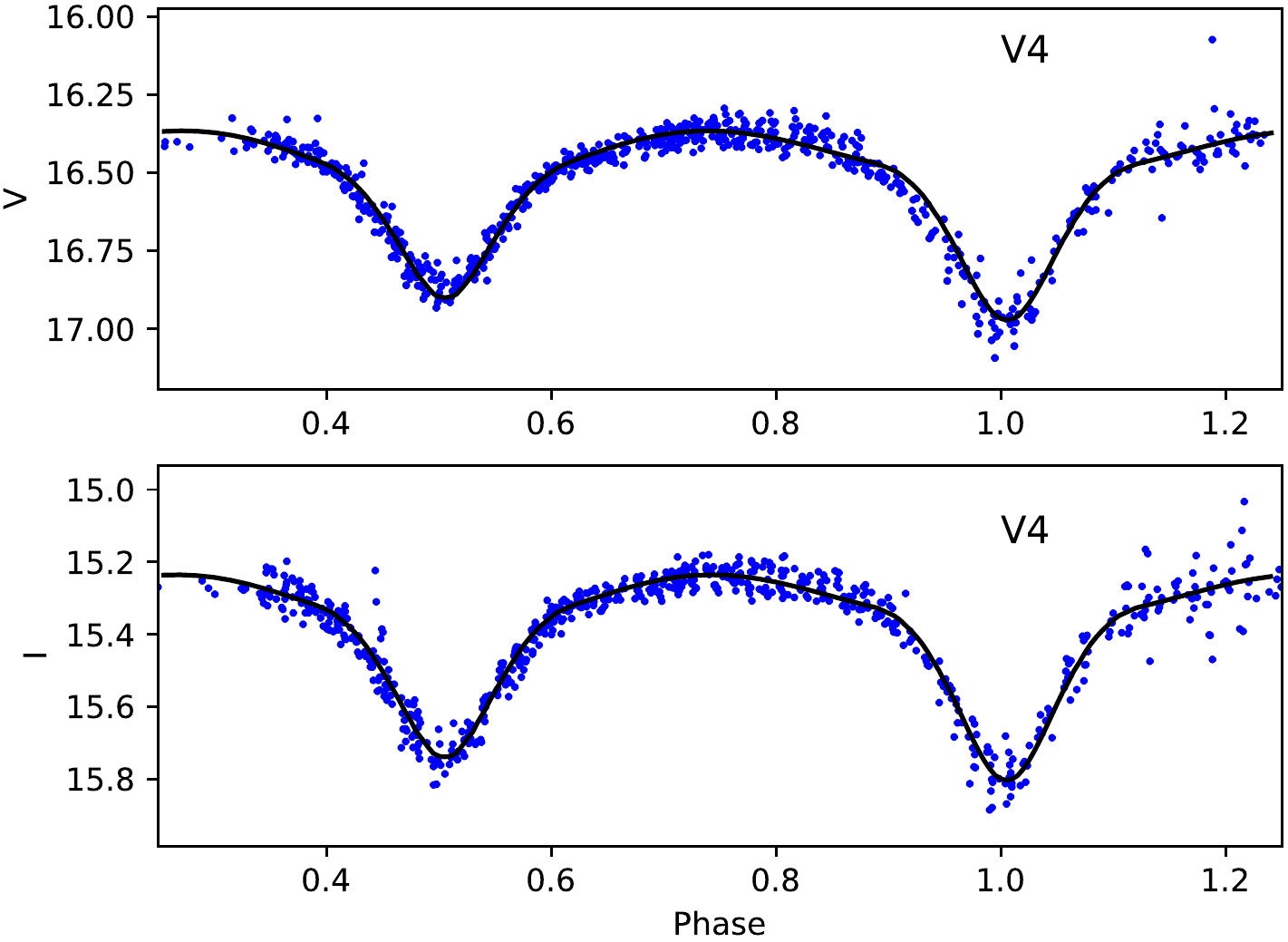}

\includegraphics[width=0.417\textwidth]{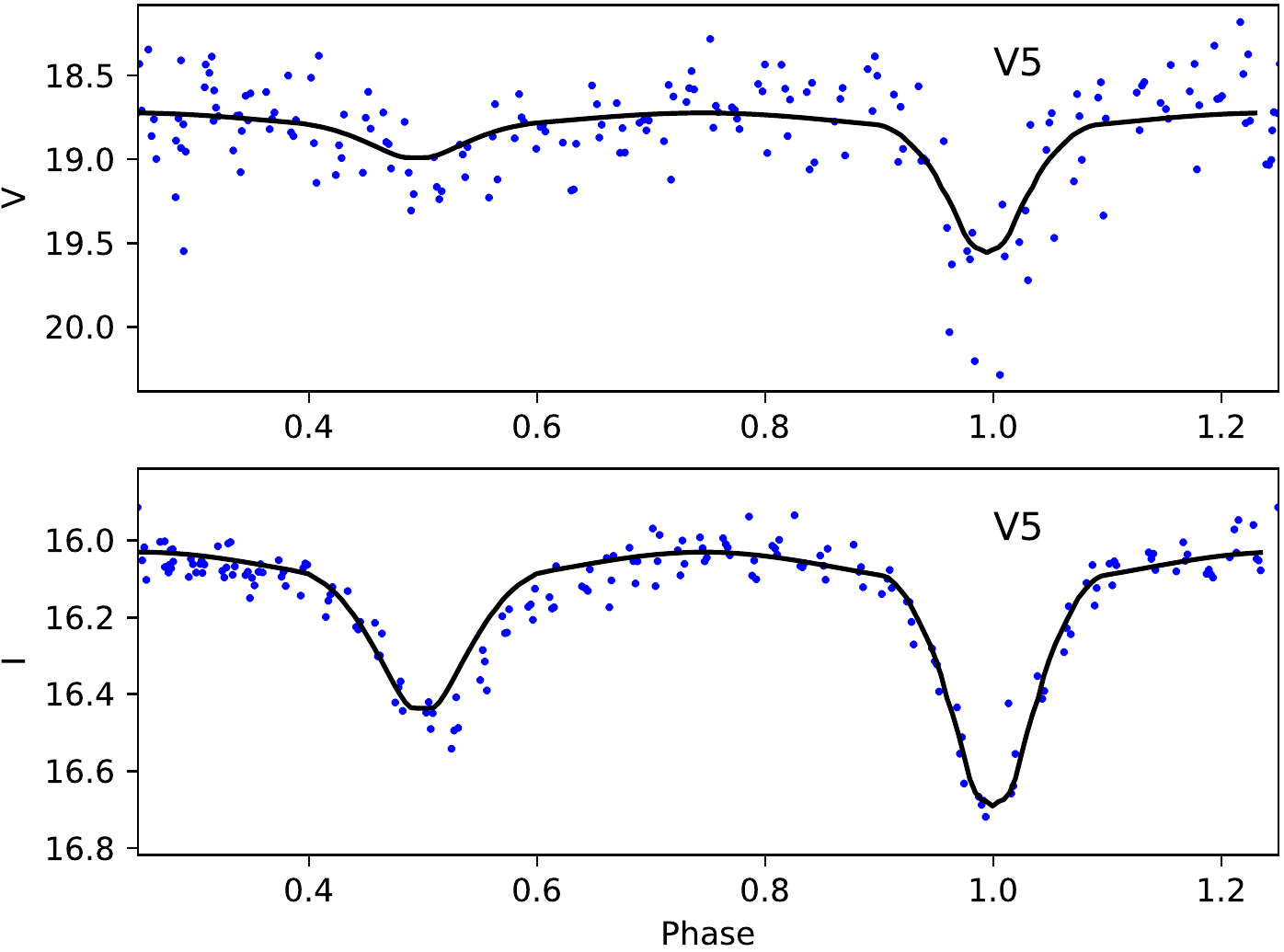}

\includegraphics[width=0.435\textwidth]{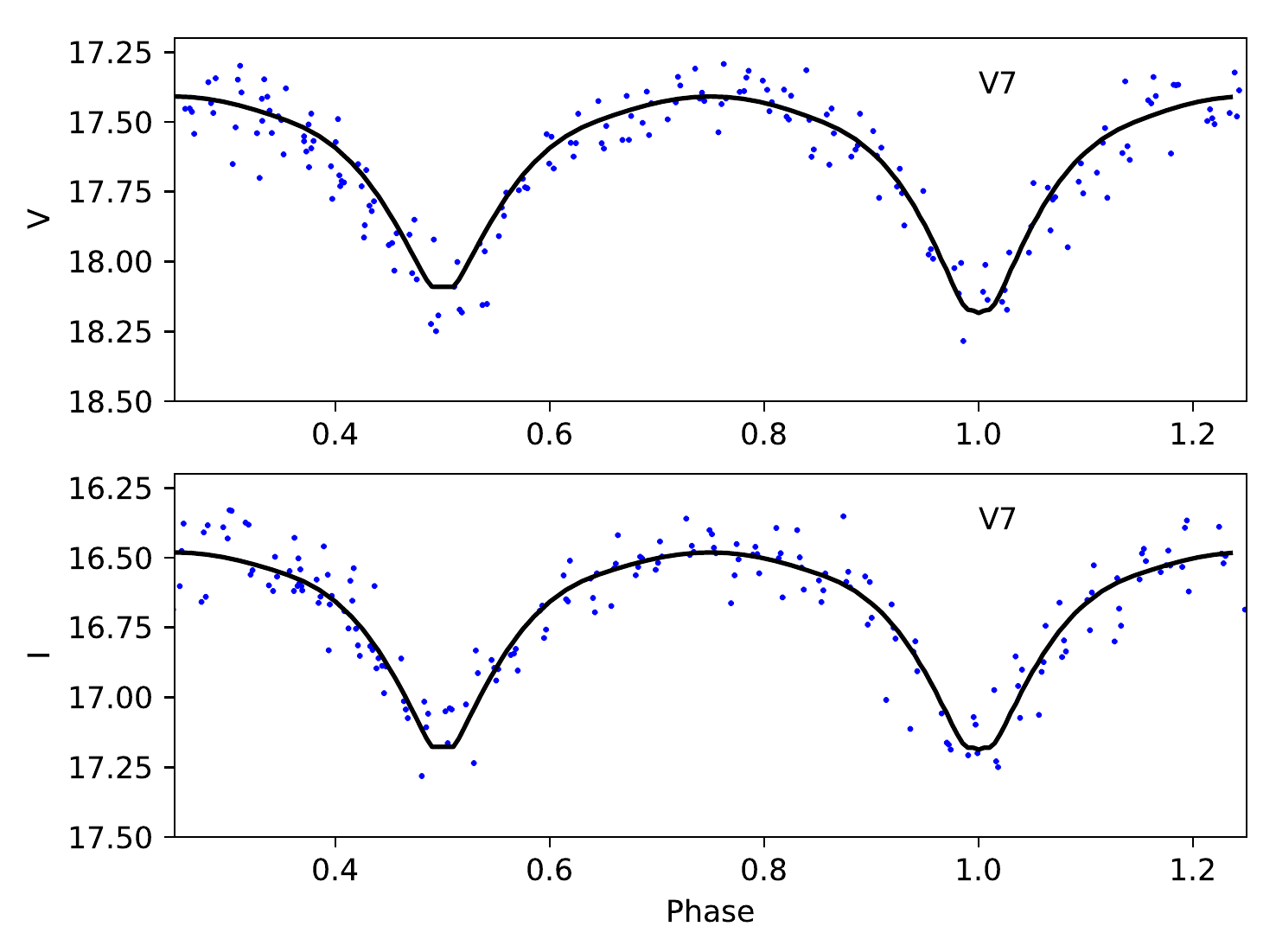}

\includegraphics[width=0.435\textwidth]{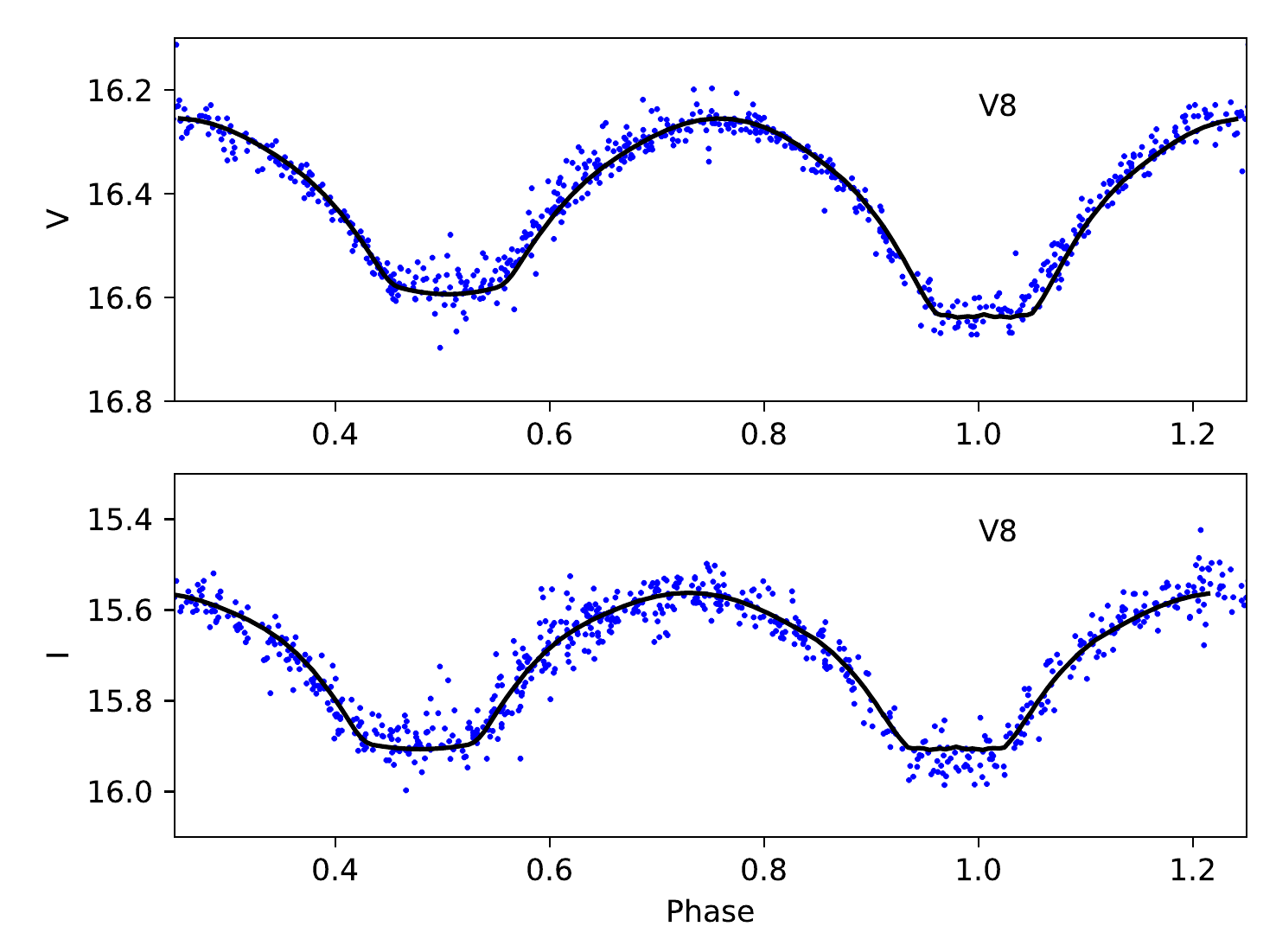}

\caption{Light curves and fits of four eclipsing binaries in NGC~6397. The  parameters of the models 
  are in Table~\ref{BinariasTab1}.   }
\label{BinariasFig2}
\end{figure}

\textbf{V4.} Assuming the star to be a cluster member, with $E(B - V)= 0.19$ and $\mathrm{[Fe/H]} = -2.0$, we obtained
$T_{\mathrm{eff,1}} \approx 5200$~K and $M_1 \approx 0.59 M_\odot$. But with these parameters,
 the calculated distance to the system turns out to be $d \approx 1290$~pc, much lower than the accepted
 distance of NGC 6397 (see \S~\ref{sec:distance}).
We therefore considered another solution, supposing that the binary is a field star of solar metallicity, i.e., with
$\mathrm{[Fe/H]} = 0.0$.  Adopting a
reasonable guess of  $E(B - V) =0.10$, we recomputed the binary parameters, which are listed in Table~\ref{BinariasTab1}, while
 the
observed and model light curves are shown in Fig.~\ref{BinariasFig2}. 
The filling 
factors $S_1$ and $S_2$ indicate that V4 is a semidetached system, with a secondary
oversized component filling its Roche lobe.
Even considering the uncertainty in $E(B-V)$, 
from the derived distance it seems very likely  that V4
  is not a cluster member.

\textbf{V5.} Our light curve in $V$ is rather faint and noisy, but the colour $(V - I)$ out of eclipse
 indicates that it is a rather red object.  Again,  starting with the assumption that it is a member of NGC~6397, 
 adopting $E(B - V)= 0.19$ and $\mathrm{[Fe/H]}= -2.0$, we derived  parameters corresponding to a very 
 low mass system, with a primary of $M_1 \approx 0.30~M_\odot$ and $T_{\mathrm{eff,1}} \approx 3700$~K,
and a distance $d \approx 550$~pc; this discards it as a member of the cluster.  We repeated the process considering
 the binary as a field star with solar metallicity and $E(B - V)= 0.05$. 
 The new results  are in Table~\ref{BinariasTab1},
and the observed and model light curves are shown in Fig.~\ref{BinariasFig2}. 
The filling factors of the model suggest a detached binary. The right side of the secondary
eclipse in $I$ looks depressed relative to the model, but due to the low quality of
the light curves we did not attempt to reproduce it. Like V4,  
it seems quite sure that V5 is also a field binary. 

\textbf{V7.} The light curves show the rounded morphology typical of  contact systems, but
the depths of the eclipses are not well defined due to  dispersion in the data. As before,
 we adopted $E(B - V)= 0.19$ and $\mathrm{[Fe/H]}= -2.0$. 
 The observed and model light curves
  are shown in Fig.~\ref{BinariasFig2}, and the parameters from the light curves fits
are  in Table~\ref{BinariasTab1}. 
The filling factors of both components indicate that V7
   is a contact system, and the resulting distance of $\approx2.6$~kpc is consistent with the assumed
   membership to NGC~6397.

\textbf{V8.} The light curves (Fig.~\ref{BinariasFig2}) present a rounded morphology with flat eclipses. 
 The  parameters derived from the modelling are in Table~\ref{BinariasTab1}, assuming  that the binary 
belongs to the cluster. 
The filling factors of both components indicate that V8
   is a contact system with a low mass ratio. The   distance ($\approx2.3$~kpc) 
makes V8 a  likely cluster member.

The binary parameters of Table~\ref{BinariasTab1} are presented without error bars,
as the unknown systematic uncertainties involved---photometry, adopted reddening and metallicity, 
 model atmosphere fluxes and limb darkening, adopted  relations for $T_{\mathrm{eff,1}}$ and $M_1$---make pointless any statistical
estimation of errors.
We can nevertheless consider, for example, the light curve of V8. By decreasing the observed $(V - I)$ at maxima in 0.02~mag 
we obtain $M_1\approx 0.96 M_\odot$,  $T_{\mathrm{eff,1}}\approx 7600$~K, and $d \approx 2480$~pc. We can therefore expect that 
uncertainties of the order of
$\delta M_1 \approx 0.03 M_\odot$, $\delta T_{\mathrm{eff,1}}\approx 200$~K, and $\delta d \approx 150$~pc would be rather
conservative. 


Additionally, we can compare the distances  in Table~\ref{BinariasTab1} with those derived
by other independent methods involving binaries. For that purpose we adopted the period-luminosity (P-L) relations of
\cite{ruc97} and \cite{ruc00} for W~UMa binaries with Hipparcos parallaxes and for W~UMa
systems in globular clusters, respectively. It must be pointed out that, as V4 and V5 are a semidetached and a detached system in our solutions,
 using  W~UMa P-L relations is not entirely appropriate;    even so, we  carried out the comparison with them as well. The metallicities used in the P-L relation of \cite{ruc97} are those in
  Table~\ref{BinariasTab1}. The distances
obtained from the P-L relations in the bands $V, I$ are listed in Table~\ref{tab:plwuma}.
We see that for the two W UMa systems, V7 and V8, the distances expected from
 the mean P-L relations are in fairly good agreement with our estimations in Table~\ref{BinariasTab1}
 and in \S~\ref{sec:distance}, and therefore with the standard distance of NGC~6397. On the
 other hand, even if they are not contact systems, V4 and V5 seem to be much nearer than the
 cluster. 

  
 

  

\begin{table*}
\caption{Parameters derived from modelling with \textsc{BinaRoche} the $V$, $I$ light curves of close  binaries in  NGC~6397}
\begin{center}
\begin{tabular}{ccccccccccccc}
\hline
Variable &  [Fe/H]$^a$ & $E(B-V)^a$  & $M_1$ & $R_1$ & $T_{\mathrm{eff,1}}$ & $S_1$ & $M_2/M_1$ & $R_2$  &  $T_\mathrm{eff,2}$
&  $S_2$  & $i$     &      $d^b$ \\ 
 & & & [$M_\odot$] & [$R_\odot$] & [K] & & & [$R_\odot$] & [K] & & [deg] & [pc] \\
\hline
   V4        & $0.00$  &  0.10      & 0.82  & 0.76  &  5000  &  0.74 &
  1.00      &  1.05     &  4875   & 1.00  & 80.5           & 1525 \\                            
V5 &  $0.00$  &  0.05  &   0.42  &  0.48  &  3475 &  0.75 &  0.46  &  0.41  &  3135 &  0.90 &  88.0 & 490 \\ 
V7 &  $-2.00$  &  0.19  & 0.68  &  \ja{0.75}  &  \ja{5950}   &  1.00 &   \ja{0.75}  &  \ja{0.66} &  \ja{5615}   &  1.00 & \ja{87.5}          & \ja{2495} \\
V8 &  $-2.00$  &  0.19  &  0.94 &  0.90    &  7420  &  1.00 & 0.32  &  0.54  &  5886 & 1.00 & 87.3 &  2350  \\
\hline
\end{tabular}
\end{center}
{\small%
\textbf{Notes.}\\
 $^a$Adopted values for the model. \\ 
$^b$Mean value of distance from $V$ and $I$.}
\label{BinariasTab1}
\end{table*} 

\begin{table}
\caption{Distances in parsecs
derived from  P-L relations for W~UMa binaries in  NGC~6397} 
\begin{center}
\begin{tabular}{cccc}
\hline
Variable & $d_V$ (RD97) & $d_I$ (RD97) & $d_V$ (R00) \\
\hline
 V4 &  1630   &   1755  &  1627  \\
  V5&  231 & 312 & 230  \\
 V7 & \ja{2354} & \ja{2383}  &  \ja{2622} \\
  V8 &  2070 &  2020 & 2305 \\
  \hline
\end{tabular}
\end{center}
{\small
\textbf{Notes.} The values $d_V$ and $d_I$ are  distances derived from light curves in $V$ and $I$.
RD97:
\citet{ruc97}; R00: \citet{ruc00}.}
\label{tab:plwuma}
\end{table}

\section{Distance to the cluster}
\label{sec:distance}

\subsection{From the SX Phoenicis P-L relation}
\label{sec:sx}

\begin{figure} 
\centering
\includegraphics[width=0.47\textwidth]{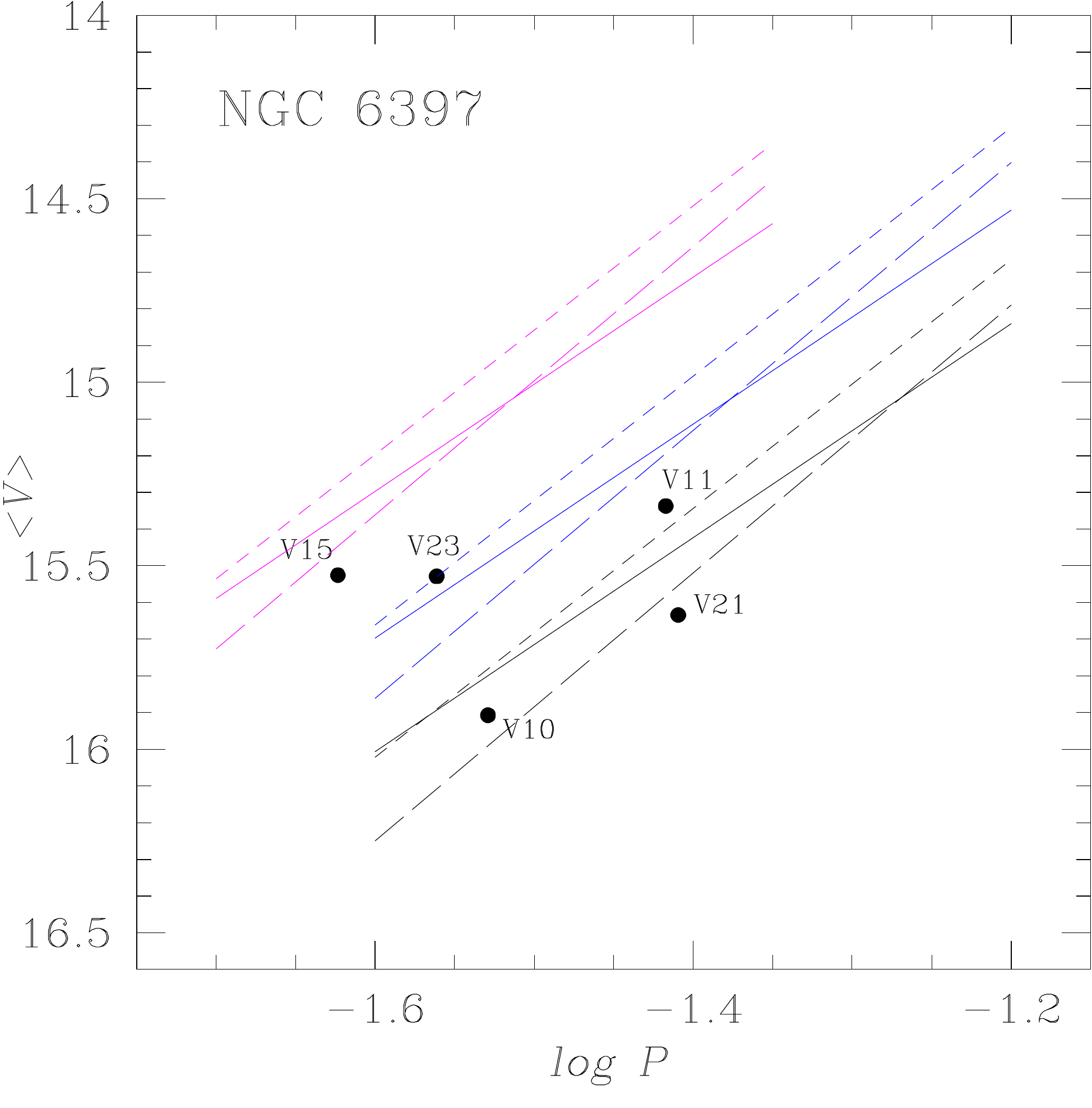}

\caption{Period-Luminosity relations for SX~Phe stars: \citet{Poretti2008} (\emph{long-dash}), \citet{Arellano2011} (\emph{solid}), and \citet{CohenSara2012} (\emph{short-dash}), for the fundamental (\emph{black}), first overtone (\emph{blue}), 
and second overtone (\emph{lilac}) modes.} 
    \label{fig:SXPL}
\end{figure}

The distance to the cluster can be estimated using the Period-Luminosity (P-L) relation for its SX~Phe stars. The variables V10, V11, V15, V21, and V23 are, given their proper motions in  \emph{Gaia}~DR2 and  positions in the CMD, very convincing SX~Phe members of the system. For the calculations we  employed three independent calibrations of the P-L relation, namely  \citet{Poretti2008} (\emph{long-dash}), \citet{Arellano2011} (\emph{solid}), and \citet{CohenSara2012} (\emph{short-dash}), as shown in Fig.~\ref{fig:SXPL}, for the fundamental (\emph{black}), first overtone (\emph{blue}), and second overtone (\emph{lilac}) modes, respectively.
The positioning of the first and second overtone loci was done 
assuming the ratios between the periods $P_1/P_0 = 0.783$ and $P_2/P_0 = 0.571$ \citep{Jeon2003,Jeon2004}. 
It is rather clear from their distribution on the P-L plane that V10, V11, and V21 are fundamental pulsators, while the main periods in V23 and V15 correspond to the first and second overtone, respectively.
Assuming these modes, and using the three calibrations for each mode, we ended up with three distance determinations for each variable---all very similar---that yield a grand average of $2.24 \pm 0.13$~kpc. This result compares  very well with the distance of 2.3~kpc in \cite{ha96},
and with the recent measurement by \cite{br18}, based on 
WFC3@HST geometrical parallaxes, of
 $2.39\pm0.10$~kpc. 

\subsection{From HB and isochrones fitting}
\label{sec:distiso}

In \S~\ref{sec:isocronas}  we showed that a
 proper matching of the Victoria isochrones to the star distribution 
in the CMD (Fig.~\ref{fig:CMD})
results by adopting a reddening $E(B-V)=0.19$ and a distance of 2.5~kpc,  in  good agreement 
with the value obtained using the SX~Phoenicis  variables in \S~\ref{sec:sx}.

 \subsection{From the modelling of eclipsing binaries}

 As  described in \S~\ref{sec:binarias}, eclipsing binaries V7 and V8 are likely cluster members; their distances, derived from the modelling of the light curves and employing P-L relations for W~UMa stars, are between 2.35 and 2.6~kpc, again consistent with the results in \S\S~\ref{sec:sx}--\ref{sec:distiso}.


\begin{figure}
\centering
\includegraphics[width=0.48\textwidth]{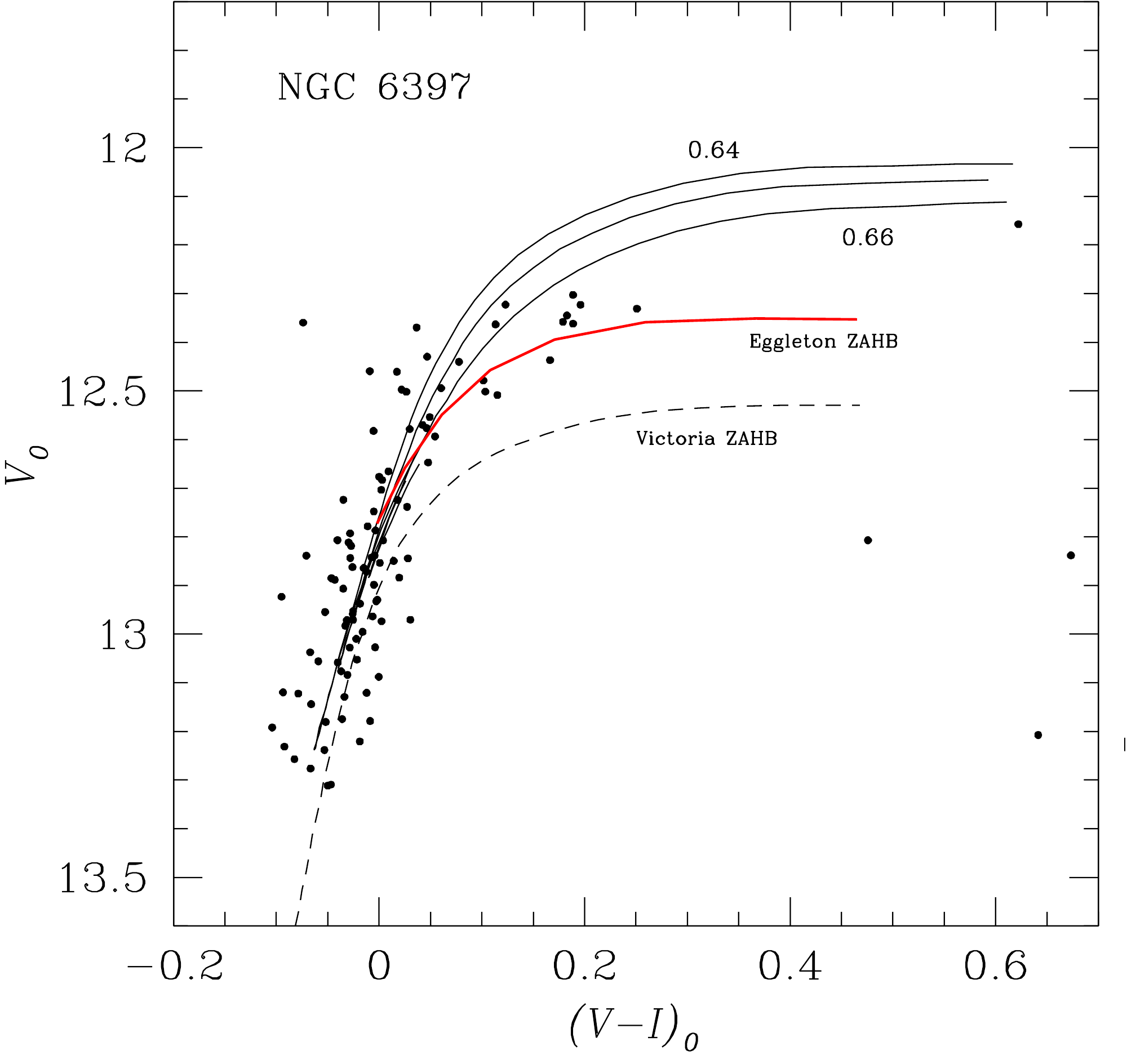}
\caption{Enlargement of the HB region in the CMD (Fig.~\ref{fig:CMD}), showing in detail the matching to the observations by the Eggleton evolutionary tracks and ZAHB, and by the Victoria ZAHB. Discussion is in \S~\ref{sec:HB}.}
    \label{fig:ZAHB}
\end{figure}

\section{The structure of the horizontal branch}
\label{sec:HB}

We now want to draw  attention to the similarity between the CMD
of NGC~6397 (Fig.~\ref{fig:CMD})  and that of   NGC~6254 (M10,  see Fig.~6 of 
\citealt{Arellano2020}). In particular, the structure of the HB is worth examining. Both clusters display a well-developed blue tail and a virtually non-existent horizontal region, with the consequent absence of  RR~Lyrae stars, with the sole exception of  the \ja{RRc} V22 in M10. The likeness extends to the 
age ($\approx$ 13~Gyr, as suggested by the turnoff isochrone fitting).

As we did for M10, we  modelled the HB of NGC~6397 with the evolution code, parametrization, and  prescription of the RGB mass-loss (a modified Reimers
law with  $\eta$ = $0.8 \times 10^{-13}$) of \citet{KPS2005}.
The code was originally
developed by Peter Eggleton
(\citeyear{Eggleton1971, Eggleton1972, Eggleton1973}),
 and further improved and 
tested by \citet[][]{Pols1997, Pols1998}  and \citet{KPS1997}.

The very blue end of the HB of NGC~6397 suggests rather
thin hydrogen shell masses, 
 consistent with the best-match HB models plotted in the CMD (Fig.~\ref{fig:CMD}, or  the magnification in Fig.~\ref{fig:ZAHB}), of total mass between 0.64 and 0.66~$M_{\odot}$. This values are similar to those we found for M10  (see  Fig.~8 of \citealt{Arellano2020}).
In Fig.~\ref{fig:ZAHB} we added the theoretical ZAHB line derived from
our Eggleton (Cambridge) code, and compared it to a standard ZAHB of the well-established Victoria models,
for [Fe/H]~$=-2.0$, [$\alpha$/Fe]~$=+0.4$, and $Y=0.25$ \citep{Vandenberg2014}.
Note that HB stars evolve upwards of their ZAHB starting points, and such a line
should, therefore, ideally demarcate a lower boundary to the distribution of  observed HB stars in the CMD.
While our ZAHB line does so, the Victoria ZAHB appears to leave a gap, indicating that the respective models
are less luminous than the observed HB stars of NGC 6397 by almost 0.2 mag.
This seems to be at odds with the fact that  both codes (Eggleton and Victoria) are mutually consistent in their results. They share the Eggleton approach of an adaptive, non-lagrangian mesh, which places more mesh points in regions with large temperature and pressure gradients.
However, the standard Victoria ZAHB uses He-core masses of approximately 0.49~$M_{\odot}$, as  D.~VandenBerg 
pointed out in a private communication. The models  we computed for M10 and this cluster, by contrast, reach a slightly higher
He-core mass of 0.51~$M_{\odot}$. This is a  small difference in mass (about 4\%) but, by virtue of the mass-luminosity relation---much like in main-sequence stars---there is
a  change in brightness 
which is fully responsible for the aforementioned
difference   of 0.2~mag.

It is pertinent to note that we systematically stop our tip-RGB evolution models just before the
He-flash, when the already slowly burning helium exceeds a luminosity of 10~$L_\odot$
\citep{Arceo2015}. Beyond this point, the exponential growth of
the He-burning leads to the He-flash and to a non-equilibrium
transition towards the ZAHB, which most evolution codes do not cover.
There is, in other words, a sort of final ``He-simmering'' time at the tip of the RGB,
especially when the hydrogen shell around the core has been much reduced  by mass loss; 
this should be expected, e.g., in old globular clusters, hence  causing the very blue HB stars
in M10 and NGC~6397. This final ``tip-RGB time'' makes the helium cores of our  models
 a little more massive than  those of the conventional Victoria ZAHB models.

\section{Summary and conclusions}
\label{sec:conclusiones}

 We have presented the photometric analysis of a new time-series of \emph{VI} CCD images
of the post-core-collapse globular cluster NGC~6397. The data reduction and the photometry
were carried out with the Differential Image Analysis (DIA) procedure. In an already 
 extensive series of papers by our group (see, e.g., AF17), we have shown 
that DIA  is a powerful tool
to generate  massive numbers of good-quality light curves for all the objects 
present in a specific sky field, with particular interest in the light curves of variable stars.
\ja{Unfortunately, NGC~6397 does not harbour RR~Lyrae  stars, which are proper for a Fourier light 
curve decomposition analysis; however, we had at our disposal another types of variables and diverse tools to take
advantage toward the determination of the cluster distance and metallicity.}


First, we cleaned the CMD making use of the powerful \emph{Gaia}~DR2 
astrometric solution and the BIRCH algorithm, selecting the most likely members
(almost 13,000 stars in Las~Campanas data, \S~\ref{membership}). After a
differential dereddening of  the CMD (\S~\ref{reddening}), isochrones for 13.0--13.5~Gyr 
and [Fe/H]~$= -2.0$ from \citet{Vandenberg2014}
were then matched to the main sequence, turnoff, and RGB (Fig.~\ref{fig:CMD}, \emph{right}).
The matching is good for
a distance of 2.5~kpc and a mean reddening $E(B-V)=0.19$.
An independent method for deriving the cluster distance involves the Period-Luminosity relation
for the
SX~Phoenicis variables present among its blue straggler population (\S~\ref{sec:sx}).
 Using three 
different P-L relations for five SX~Phe stars (V10, V11, V15, V21, and V23),
 we obtained a mean value of  $2.24\pm0.13$~kpc.
Yet another distance determinations come from the cluster's
eclipsing binaries V4, V5, V7, and V8. A first method uses the modelling of their light
curves with the \textsc{BinaRoche} code (\S~\ref{sec:binarias}); it turns out that, while V4 and V5 are a semidetached
and a detached system fairly nearer than the cluster, V7 and V8 are
contact binaries at 2.6 and 2.3~kpc, thus being  likely cluster members 
(Table~\ref{BinariasTab1}). These  
results are furthermore confirmed by  using a P-L
relation for W~UMa binaries (Table~\ref{tab:plwuma}).

The blue horizontal branch of NGC~6397 is similar to that of M10 which, however, harbours just one
RR~Lyrae   \citep{Arellano2020}, while NGC~6397 has none. The models 
of 0.64--0.66~$M_{\odot}$ with mass loss at the RGB \citep{KPS2005}, located at the distance and reddening
suggested by the isochrones fitting to the TO and RGB, provide a good matching of the HB stars. Comparing
with M10, this means that a minor range of mass loss at the RGB---i.e., a minor range
of mass of the remaining envelopes above de He-core---represents better the observations.
By contrast, in M10 that range is wider, 0.56--0.62~$M_\odot$; it must be pointed out, however,
that the HB of M10 is richer and extends
  longer towards faint magnitudes than that of NGC~6397. The difference of mass loss
	during the RGB stage might be abscribed to magnetic fields in the stars'
	chromospheres that somehow hinder the normal losses \citep{Arellano2020}.

\section*{Acknowledgements}

AAF acknowledges the support from DGAPA-UNAM grant through project IG100620. 
It is a pleasure to thank Dr.\ Don A.\ VandenBerg for the use of his model interpolation software and for an instructive discussion on  horizontal branch models.



\bibliography{bibliografia}

\end{document}